\documentclass[]{mn2e}

\usepackage{amssymb}
\usepackage{graphicx}
\usepackage{amsmath}
\usepackage{graphics}

\usepackage{color}

\title[Magnetic dust polarizers] {Candidate magnetic dust structures for star light polarization}

\author[R. Papoular]{R. Papoular$^{1}$\thanks{E-mail:papoular@wanadoo.fr}\\
$^{1}$Service d'Astrophysique and Service de Chimie Moleculaire,\\
CEA Saclay, 91191 Gif-s-Yvette, France}

\begin{document}

   \maketitle
\label{firstpage}

\begin{abstract}
 Rotation damping and alignment are discussed as prerequisites for polarization. An expression is derived from first principles, for the damping time of the rotation of a particle in a magnetic field, under the Faraday braking torque, provided its electrical properties are known. This makes it possible to describe mathematically, in great detail, the motion of the particle and determine its ultimate state of motion, if a steady state is possible at all. This work defines, first, the necessary condition for the Faraday braking to be effective: a) the net electronic charge distribution should not be uniform throughout; b) the number of vibration modes should exceed a few tens. Resonance of rotation frequency with any of these modes is not a requirement. For alignment to be possible, the ratio of gyroscopic and conservative magnetic to non-conservative (retarding) magnetic torques must be low. Either dia-, para- or ferro-magnetism can do, and a small susceptibility is enough and even preferable. This opens up a wide spectrum of possible candidates. A few examples are given.

\end{abstract}


\section{Introduction}

Soon after the polarization of starlight was first measured, it was suggested that ferro-magnetic needle-like grains could be responsible for this phenomenon (see Spitzer and Schatzman 1949, Spitzer and Tukey 1951): such grains would align with the magnetic field like the needle of a terrestrial compass. Subsequently, Davis and Greenstein \cite{dg} objected that, because of interaction with ambient hydrogen atoms, each grain would be endowed with a non-negligible angular moment, in which case the gyroscopic torques would overcome the magnetic torque due to any realistic combination of magnetic field intensity and grain magnetic moment, sending the grain in precession and nutation around the magnetic field. They went on to suggest that this could be remedied by a sufficiently strong braking torque, as is the case for a heavy spinning top, due to friction at the pivot. They suggested, moreover, that such braking could be implemented by giving a paramagnetic  grain sufficient angular velocity that the torque due to the imaginary component of the susceptibility becomes dominant. Purcell \cite{pur79} later went to great lengths to increase that initial angular velocity to supra-thermal values, up to $10^{9}$ rad/s, noting that the Barnett effect would then strongly contribute to alignment due to the increase of the apparent angular velocity of the field in the grain's frame. This has since been one of the dominant paradigms in the treatment of alignment.

Now, a new braking process was recently demonstrated to be remarkably strong even in the absence of particularly high imaginary magnetic susceptibilities (Papoular 2017; referred to as P17 below); this becomes apparent when the internal structure of the grain is taken into account. One then finds that, even if the grain is electrically neutral as a whole, the density of electrical charge varies from atom to atom: this is quantified by associating an equivalent positive or negative (``Mulliken'') net charge, $q$, to each atom. Each of these, as part of the general grain motion, also has a velocity, $v$. In a field $\vec H$, each atom then feels a Faraday force, $q\vec v\times\vec H$. In this paper, I show how this gives rise to a braking torque, independent of the grain magnetism, which may be computed from first principles, given the elementary structure of the grain and with the help of chemical modeling.

The fact that the braking torque may be computed from first principles allows one to lay down the explicit equations of motion of the grain and solve them numerically. By contrast with usual treatments of alignment, one can then follow the process all along. This is done here for 4 examples: a diamagnetic and a ferromagnetic molecule, as well as a bigger para- or ferro-magnetic grain.

These examples show that supra-thermal angular velocities are no longer a requirement for alignment. A useful byproduct of this procedure is to highlight the importance of the relative strengths of the gyroscopic, magnetic and braking torques acting on the same grain, which, in turn, helps defining desirable profiles for candidate polarizers: magnetic properties, shape, etc.

\bf Light polarization\rm by a molecule depends not only on its alignment but also on the intrinsic polarization of its natural vibrations, and so requires a specific computation, an example of which is also given. Since the internal structure of a big grain is generally disordered, it requires a different treatment, an example of which is given too.\rm

The paper is organized as follows. 

-Sec. 2 : rotation damping, itself subdivided into 2.1: expression for the braking time; 2.2: necessary conditions for damping (non-homogeneous distribution of net electrical charges; large number of vibration modes); 2.3: the physical essence of Faraday braking; relation to the fluctuation-dissipation theorem;

-Sec. 3. 3.1: dynamics in a magnetic field: the equations; alignment requires low relative strengths of gyroscopic and conservative torques; examples: 3.2: a diamagnetic molecule; 3.3: a ferro-magnetic molecule; 3.4: a big graphitic grain; 3.5: time constraint on alignment: steady rotation state (or complete damping) must be reached in between two rocketing atomic impacts;
 
-Sec 4 : polarization is treated next, using examples of possible candidate polarizers.

\section{Rotation damping}

\subsection{The braking time of the Faraday mechanism}
Chemical modeling ``experiments'' complemented by a mathematical model have recently demonstrated the existence of a rotation braking torque associated with the Faraday force, $q\vec V\times\vec B$, acting upon the positive and negative partial (Mulliken) atomic charges of any electrically neutral diamagnetic dust paricle carrying an electric dipole moment (Papoular 2016) . We recall that the atomic partial charges are a measure of the spatial electric density distribution throughout the otherwise neutral particle. The dipole moment results from the separation of the centers of the positive and negative partial charges. In this mechanism, the energy lost by the slowing rotation is absorbed by the particle natural vibrations. Regarding this braking process, many issues have now to be addressed, in particular relating to the physical understanding of this mechanism and to its possible contribution to the polarization of starlight by interstellar dust. The present work addresses the following among these issues:

-The need for as accurate an expression of the braking time, $\tau_{b}$, as possible;

-What the necessary conditions are for Faraday rotation braking;

-Why this mechanism does not require resonance between the rotation frequency and a normal vibration frequency of the particle;

-Why is magnetism required although the corresponding susceptibility does not enter the braking time;

-What is the impact of the various parameters on the ultimate orientation of the molecule;

-The determination of $\tau_{b}$ for candidate dust particles.

Papoular \cite{pap16} offered a grossly approximate expression for $\tau_{b}$, featuring the dipole moment, $D$, and the radius of gyration, $R_{G}$, corresponding to the rotation axis of interest. A much more accurate expression can be obtained as follows, using the capabilities of the same Hyperchem chemical modeling package, \bf available from Hypercube Inc., publication HC50-00-02-00; for details, see Papoular \cite{pap16}.\rm

\begin{figure}
\resizebox{\hsize}{!}{\includegraphics{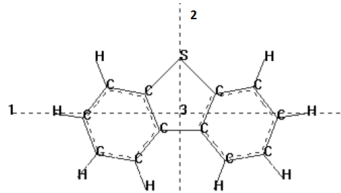}}
\caption[]{The small, carbon-rich, diamagnetic molecule, dubbed trioS, used in the chemical modeling experiment of P17. Principal axes 1,2,3 of the molecule (dashed lines) are set along figure axes Ox, Oy, Oz  respectively.}
\label{Fig:trioS}
\end{figure}

Consider the same trioS molecule as in Fig. 1 of P17, and in Fig. \ref{Fig:trioS} here, with the magnetic field, $\vec H$, parallel to axis y of this figure. Let the particle rotate about its principal axis 1, initially coinciding with axis Ox of the figure, as in the chemical modeling experiment studied in detail in that paper and taken here as an example. The package delivers the position of each atom, $\vec r(xyz)_{i}$, with the center of mass as origin, together with its partial (Mulliken) charge, $q_{i}$. The center of charge of a group of charges is $\Sigma (q_{i}\vec r_{i})/\Sigma q_{i}$. In the present case, the sums of positive and negative charges are + and -1.11 electron charges (1 e=$4.8\,10^{-10}$ statCoulomb), and their  centers of charges, $\vec r_{+}$ and $\vec r_{-}$, respectively, fall along principal axis 2 (parallel to Oy), at $y$=-0.202 and -0.159 \AA{\ }, respectively.

 Each positive or negative partial atomic charge is subject to the Faraday force, $q\vec V\times\vec B$. The coupling between molecular bulk rotation and internal vibrations is ensured by the presence, in this expression, of the total charge velocity, $V$, which is the vector sum of the local rotation velocity \it and all the velocities imparted to the charge by the superimposed normal vibration modes of the molecule.\rm This fulfills one of the necessary conditions of  rotation braking. \it The second necessary condition requires that the conversion of rotation to vibration energy be irreversible.\rm Section 2.3 below shows that this is all the more fulfilled that the number of vibrations increases. 

In the limit where the number of vibrations is large, it is not possible to express $V$ rigorously. As in Papoular \cite{pap16}, assume therefore, as a mathematical approximation, that, \it for each vibrating charge, the number of superimposed normal modes is so large that the vector sum of all corresponding velocities averages out, leaving only the local angular velocity.\rm The sum total of all torques on the partial charges is then

\begin{equation}
\vec\Gamma=\Sigma q_{i}\vec r_{i}\times(\vec r_{i}\times\vec \omega)\times\vec H/c,
\end{equation}

in the Gauss escgs system of units, in which both $B$ and $H$ are expressed in Gauss and equal in magnitude. This reduces, here, to $<H>\omega\Sigma q_{i}y_{i}^2/c$, where $c=3\,10^{10}$ cm/s is the velocity of light, and $<H>$ designates the average magnitude, over a rotation period, of the component of $\vec H$ lying parallel to molecular axis 2.  On conditions to be discussed in next section, the work done by this torque is entirely spent in molecular vibration energy, at the expense of rotation energy. Let $I$ be the moment of inertia for the selected rotation, and assume no other force acts upon the molecule; then, the equation of motion reduces to $I\dot\vec\omega=-\vec\Gamma$. This gives a damped rotation with (braking) time

\begin{equation}
\tau_{b}=-\frac{\omega}{\dot\omega}=\mid\frac{Ic}{<H>\Sigma (q_{i}y_{i}^2)}\mid.
\end{equation}

Remarkably, \it  $\tau_{b}$ scales exactly with $H^{-1}$, and the magnetic susceptibility is absent from the expression.\rm In the example considered here, $I=5.4\,10^{-38}$ g.cm$^{2}$, $H= 8.5\,10^{8}$ G, \bf as in P17  \rm; $\Sigma (q_{i}y_{i}^2)=3.37$ e$\AA{\ }^{2}$ is the newly computed braking parameter \rm and $<H>\sim0.6\,H$ (average over a full rotation), so $\tau_{b}=20$ ps. This value is indeed close to the value deduced from the chemical experiment, in P17: 27 ps, \bf comforting the relevance of this treatment to the physics involved. \rm

It must be recalled that formula 2 was derived above for a very particular configuration: rotation about principal axis 1 in a field parallel to axis 2. In general, if the field makes an angle $\theta$ with the rotation vector, only its component normal to the rotation is active, so in eq. 2, $H$ must be replaced with  $H$sin$\theta$, which is true, in particular, for $\vec H$ pointing in the Oz direction of Fig. \ref{Fig:trioS}. The same applies to the other 2 principal axes, showing that rotation about the magnetic field direction is not damped. Also, in general, the quantity $y$ in eq. 2 must be replaced by $r$, the distance of the charge to the rotation axis, for it does not always coincide with one of the coordinates. Finally, eq. 2 becomes

\begin{equation}
\tau_{b}=-\frac{\omega}{\dot\omega}=\mid\frac{Ic}{<H>\mathrm{sin}\theta\Sigma (q_{i}r_{i}^2)}\mid.
 \end{equation}
 
 In the case of our plane trioS molecule, this result must now be applied to rotation about principal axes 2 and 3. For the former, $r=x$, so $\Sigma (q_{i}r_{i}^2)=3.72$  e$\AA{\ }^{2}$. For the latter, $r^{2}=x^{2}+y^{2}$, so  $\Sigma (q_{i}r_{i}^2)=7.1$ e$\AA{\ }^{2}$. If  $\Sigma (q_{i}r_{i}^2)$ is taken as a measure of the relative efficiency of the braking, it appears, in the case of the trioS molecule, that it is of the same order of magnitude for all 3 principal rotation directions. In P17, the empirical efficiency was measured by the quantity $s$, (dimension QLT in the escgs system) and found by chemical modeling to be $8\,10^{-36}$ escgs. It can now be expressed $ a \,priori$ by 
 
\begin{equation}
s_{j}=\Sigma_{i} (q_{i}r_{i}^2)/c\,.
\end{equation}

For rotation about principal axis 1, here, this amounts to $5.5\,10^{-36}$, in good agreement with chemical modeling. \bf Although, for practical computation ease, the magnetic field intensity assumed in this modeling experiment is orders of magnitude higher than interstellar fields, this agreement lends credence to the concept of a braking torque due to a Faraday force. Its quantitative expression through $s$ allows one to study the dynamics of a particle in any field (Sec. 3).\rm 

\bf If an external electric charge is deposited locally onto the particle, it will promptly be redistributed all over the latter according to the spatial distribution of atoms. The chemical algorithm automatically computes the new distribution of net charges according to quantum mechanics; expression (4) must be recomputed, but this, in general, only amounts to a quantitative change of the dynamics.\rm
 
 In the following, for clarity, we will refer to $\tau_{b}$ as the quantity $I/(sH)$, remembering that this is smaller than the true value given in eq. 3, by a factor of order 1 resulting from the time variations of $H$ and $\theta$. For given particle composition and structure, $I\propto a^{5}$ and $s\propto a^{3}$, where $a$ is the linear dimension of the particle. Consequently, $\tau_{b}\propto a^{2}/H$. This differs from the Davis-Greenstein damping time by the power 1 of $H$, instead of 2, and the absence of the temperature.

While $\tau_{b}$ is a precious information regarding alignment, it does not alone tell the whole story. This is because the molecule is also subject to exchanges of rotation between the 3 principal axes, as well as torques formed by the magnetic field with the molecular magnetic moments. This blurs the picture of a given rotation continuously damped by whatever conversion mechanism and may even interrupt the latter. To figure out the whole process, and determine its total length in time, one can have recourse to the numerical solution of the relatively simple system of governing equations (as, for instance, in Papoular 2017), since the parameters which appear in these equations are now all available, at least qualitatively. This is done in Sec. 3.

 \subsection{Necessary condition for Faraday braking}
 
 Not all structures can be subject to Faraday braking. A necessary condition is immediately apparent  in eq.2: $ \Sigma (q_{i}r_{i}^2)$ must be finite. This does not imply that the particle must be charged, for the $q$'s designate partial (Mulliken) charges (which add up to zero, by definition) as well as externally added arbitrary charges. \it It only means that the electronic charge distribution should not be uniform throughout \rm: a homogeneous and perfectly symmetric structure will not do. While the presence of a finite dipole moment implies that Faraday braking applies, this is not a necessary condition. Thus, a circular disc with charges of different signs uniformly distributed over different rings is acceptable even though it has no dipole moment.

 In the limiting case of large particles, one can hardly imagine dust grains to be geometrically symmetric and chemically homogeneous structures. They are most probably aspherical and irregular in shape. In classical treatments of polarization (cf. Spitzer 1978), large particles are assumed to be strongly prolate or oblate ellipsoids like discs or rods, to provide polarization  and sometimes even alignment. Regarding Faraday braking, this is not a requirement: it is enough that the structure be highly disordered, so as to carry local,randomly oriented, small dipole moments, and satisfy the necessary condition of non-uniform charge distribution, just defined. For instance, a spherical aggregate of small graphitic ``bricks'' will be subject to braking.
 
 As for polarization, it is essential to realize that, if the particle is not made of a microscopically homogeneous material, it has different polarization capability in different directions, even if $H=0$; this is because the different normal modes of vibration have different polarization directions and intensities. As a consequence, unless the particle is perfectly spherical and, at the same time, made of a homogeneous material, it will have some finite polarizing capability. The calculation procedure and result was given for the trioS molecule as an example in Sec. 2.2 of P17.
 
 The picture that emerges suggests that the range of unacceptable candidates for our present purposes is very limited indeed. It includes only highly improbable shapes and structures. In this context, it is hard to figure out general polarization rules. Each case is a special case and requires its own computation or modeling (or measurement). A few examples are given below.
 
 \subsection{Why resonance is not a requirement for Faraday braking}

By resonance is meant, in this title, the equality of the rotation frequency with one of the natural frequencies of oscillation of the structure. In fact, resonance between rotation and vibrations is automatically satisfied due to the presence, in the \it resultant \rm velocity, $\vec V$, of each atom of a \it component \rm from each normal vibration frequency (Sec. 2.1).

 In the hypothetical limit, where the particle has only one vibration mode, the situation is similar to that of two coupled resonant circuits. (The coupling, here, is provided by Faraday's  force, whose expression features the velocity of the vibrating charge). Then, either they are out of resonance and energy flows forwards and backwards from one to the other and in very small amounts, or they are tuned to resonance and energy begins to flow at a constant rate from rotation to vibration; however, they soon fall out of resonance as the rotation frequency decreases along with its energy and energy conversion stops.

By contrast, if the molecule includes a large number of vibrating charges, and its vibration spectrum is considerably extended, energy flows from one vibration to the other (as they are all more or less coupled). Then, the sum total of all Faraday forces only depends on the rotation velocity, as the sum of vibration velocities averages to zero. In other words, the reaction of the vibrations on the rotation is neutralized so vibration energy cannot flow back to rotation and energy conversion goes on to exhaustion. In the electric circuit analogy, this corresponds to a resonant circuit feeding energy into an infinitely long electric line, or a matched line (i.e. terminated by its characteristic impedance). This is the second necessary condition for any braking mechanism: \it that the conversion of rotation to vibration energy be irreversible. \rm It is a condition for the expression (3) of $\tau_{b}$ to be valid.

This argument implies that the rotation frequency need not be resonant with any of the vibration frequencies. This is true even at very low rotation speed as the coupling of rotation and vibration always applies. The chemical experiment reported previously (P17) suggests that this is still true for as small a molecule as one bearing only 21 atoms (57 vibration modes).

It is also remarkable that the molecule need not be vibrating beforehand for the conversion process to start, as the Faraday coupling between rotation and vibrations is already active through the rotation velocity. 

The Faraday braking effect is similar to the friction that is invoked to explain the heat generated by hysteresis when a ferromagnetic material is subjected to a variable magnetic field. In that case, friction occurs between randomly magnetized domains trying to align with the external magnetic field. In Faraday braking, friction is between electrons and atoms.

This brings to mind another analogy: that with a macroscopic projectile moving through a gas as treated by the Langevin equation. Braking of its velocity will occur even if the gas temperature is very low, provided its density is finite as is the collision cross-section offered to the projectile. With this analogy, the dual pairs are velocity/rotation speed, mass/inertia, gas molecules/vibration modes, collision cross-section/Faraday force, friction coefficient/$\tau_{b}^{-1}$.

Classically, the Langevin equation and the associated fluctuation-dissipation theorem are usually demonstrated using the model of a projectile moving in a gas assuming near thermodynamic equilibrium and an infinitely extended uniform spectrum of fluctuations ( Reif 1984). Kubo \cite{kub} demonstrated a generalized form of the theorem, which extends to irreversible phenomena and non-white fluctuation spectra. Accordingly, even though thermodynamic equilibrium does not obtain in our case, the essence of the theorem still applies here: molecular vibrations absorb friction heat and, at the same time, are the very condition of this friction.

\begin{figure}
\resizebox{\hsize}{!}{\includegraphics{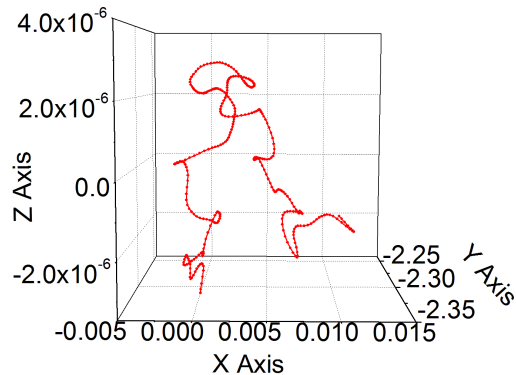}}
\caption[]{A chemical modeling experiment illustrating a form of the fluctuation-dissipation theorem (Johnson-Nyquist noise): a trioS molecule \it at rest \rm at T=0 K is immersed in a field $H=8.5\,10^{8}$ G. The graph displays a partial trajectory of the positive tip of the molecular electric dipole moment (average value in the magnetic field: 2.3 Debye).  The coordinates are its 3 components in lab space (in Debyes). Total time: 0.34 ps, elementary calculation step: 0.001 ps. This reveals a nearly random walk superimposed on pendulum oscillations about the magnetic field (parallel to Oy).}
\label{Fig:rndwalk}
\end{figure}

\begin{figure}
\resizebox{\hsize}{!}{\includegraphics{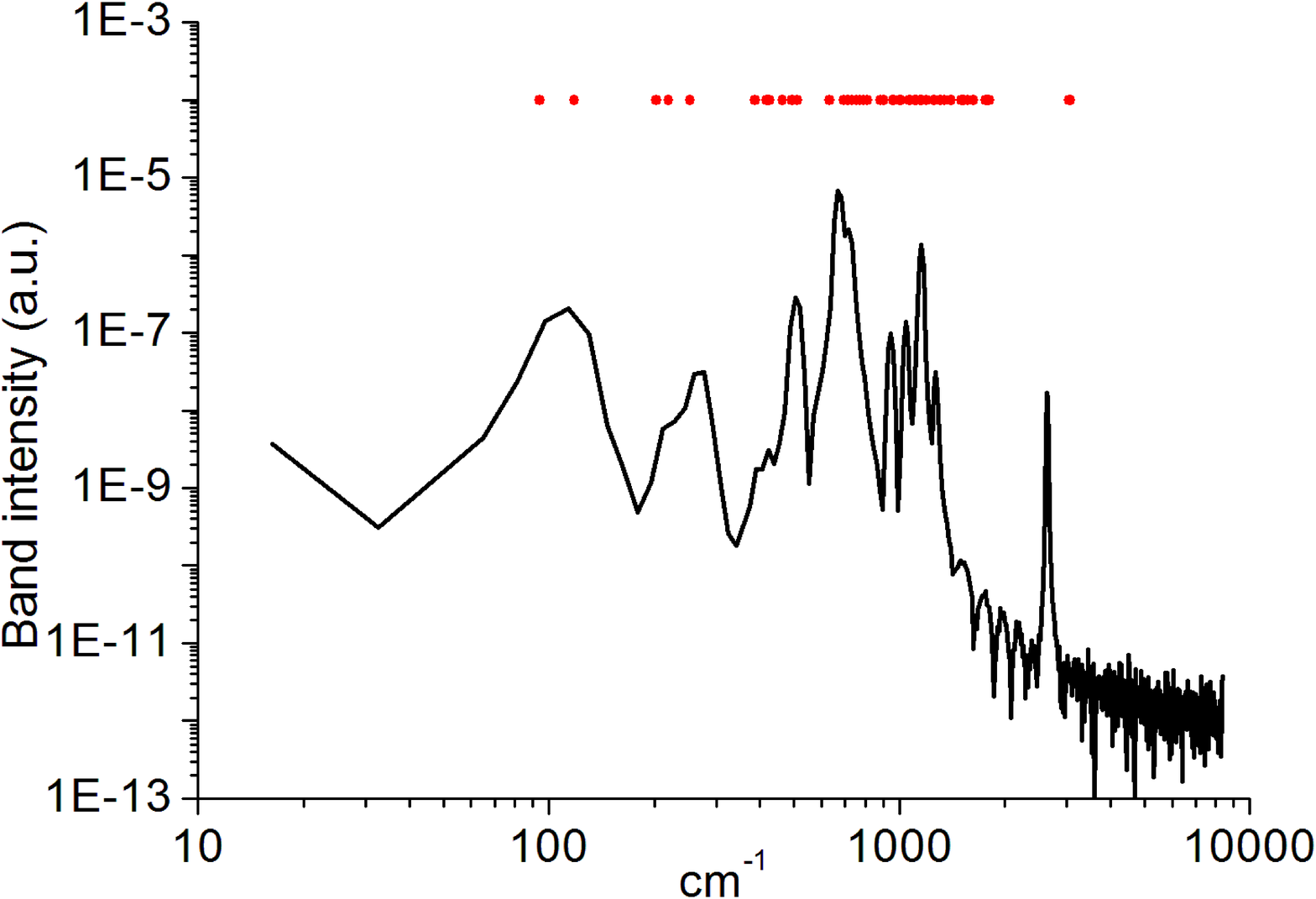}}
\caption[]{The Fourier Transform of the auto-correlation function of the dipole moment magnitude in the chemical experiment of Fig. \ref{Fig:rndwalk}. It displays a series of bands extending over the spectral range of the molecular normal vibrations (indicated by dots with ordinate $10^{-4}$); this demonstrates the coupling, through the Faraday force,  between bulk motion and internal vibrations of the molecule, as expected from the mechanism imagined in this work. Because of the requirement of high time resolution, the overall time length of this experiment was only 0.34 ps; whence the broadening of the vibration lines and the absence of most pendulum (low frequency) oscillation components.}
\end{figure}
\label{Fig:chargespec}

Another form of the the fluctuation-dissipation theorem is Nyquist theorem and predicts the Johnson-Nyquist noise in pure electric conductors, due to the fluctuation of conducting electrons. A similar phenomenon was demonstrated using the same Hyperchem package as in Sec. 1. Starting with the trioS molecule at rest at 0 K, the sudden application of the magnetic field (here, along the Oy axis of Fig. \ref{Fig:trioS}) rocks the charges, whose ensuing motion is then reflected in the variations of the 3 components of the total dipole moment (Fig. \ref{Fig:rndwalk}). Its main features are pendulum oscillations about the magnetic field (due to the torques formed by the field and the 3 components of the induced molecular magnetic moments) and random excursions superposed upon them. The spectral content of the random motion is revealed by the Fourier Transform analysis of the auto-correlation function of the dipole moment magnitude (Fig. \ref{Fig:chargespec}). This clearly displays a series of bands extending over the spectral range of the molecular normal vibrations (indicated by dots with ordinate $10^{-4}$), as expected from the mechanism imagined in this work. Because of the requirement of high time resolution, the overall time length of this experiment was only 0.34 ps; whence the broadening of the vibration lines and the absence of most pendulum (low frequency) oscillation components. It is clear, however, that the spectrum is far from being the ``white'' spectrum required by the Langevin equation. In agreement with Kubo's generalized theorem, this experiment shows that even such a small molecule, with its sparse vibration spectrum, may be subject to Faraday rotation braking, albeit, perhaps to a lesser extent.

Still another way to look at the Faraday braking is to consider that the rotating magnetic field periodically perturbs the potential wells in which the electrons oscillate. As a consequence, the electronic vibrations are modulated, giving rise to vibration side-bands. The energy of the Stokes phonons is exchanged between vibrations as the action of all the asynchronous vibrations cannot recombine into rotation. The energy of the anti-Stokes phonons accumulates in the mother vibrations. The larger the number of vibrations, the more complete is the conversion.

 \section{Alignment in a magnetic field}
 We are now in a position where the equations of motion of a particle can be written from first principles and excluding simplifying assumptions, provided its mechanical, electrical and magnetic properties are known and the initial conditions defined. 
 
\subsection{Dynamics in a magnetic field}

In the particle's frame of reference, the orientation of the particle is governed by the vectorial relation (Goldstein 1980)
\begin{equation}
\dot\vec L+\vec\omega\times\vec L=\vec N\, ,
\end{equation}

where $\vec L, \vec \omega, \vec N$ are, respectively, the particle's angular momentum and velocity, and total external torque acting on it.

Let $I_{i}$, where  i=1,2,3, be the 3 principal moments of inertia of the particle, and $\omega_{i}$, the corresponding rotation speeds. \it In the reference frame of the molecule \rm, $L_{i}=I_{i}\omega_{i}$ and eq. 5 can then be developed into the following set of equations, \bf in the Gauss system of units\rm :

\begin{eqnarray}
I_{1}\dot\omega_{1}-\omega_{2}\omega_{3}(I_{2}-I_{3})=N_{1}-s_{1}H\mid\mathrm{sin}p\mid\omega_{1}\\
I_{2}\dot\omega_{2}-\omega_{3}\omega_{1}(I_{3}-I_{1})=N_{2}-s_{2}H\mid\mathrm{sin}q\mid\omega_{2}\\
I_{3}\dot\omega_{3}-\omega_{1}\omega_{2}(I_{1}-I_{2})=N_{3}-s_{3}H\mid\mathrm{sin}r\mid\omega_{3}\, ,
\end{eqnarray}

where the second, third and fourth terms from the left of each equation are, respectively the gyroscopic, external and braking torques. Here, $\vec N=\vec M\times\vec B$, is the torque formed by the field and the magnetic moment it induces in the molecule, $M_{i}=\chi_{i}VH_{i}$, where $\chi$ is the susceptibility and $V$, the overall volume. $s_{i}$ is no longer an empirical parameter but is now given by eq. 4 above, as a function of the charge spatial distribution (even though the particle may be globally neutral). Finally, $p,q,r$ are the angles between the field and the 3 principal axes, respectively. In the Gauss system, magnetic induction $B$, and field, $H$, are expressed in Gauss and Oersted, respectively. Since their numerical values are the same in this system, the term ``field'' will be used for both and given in Gauss. 

The details of the numerical computations are here the same as in P17, except for a sign error in the expression of $\omega_{3}$ in their eq. 7. The opposite sign is adopted here, in conformity with the conventional direction of the angular velocity vector. 

Because these equations are non-linear in $\omega$ and $H$, it is not easy to find simplifying assumptions that would allow solutions to be computed analytically. Instead, numerical computations have to be made for a few combinations of parameters that would hopefully represent typical dust particles in particular interstellar environments. For instance, carbon-rich dust will usually be diamagnetic, with different negative susceptibilities in the 3 principal directions of the particle. On the other hand, the electrons of various atomic impurities (metals, in particular) may impart paramagnetism to otherwise non-magnetic material (Kittel 1956); in this case, all 3 susceptibilities are positive and, likely, equal. Finally, iron atoms carry a spontaneous magnetic moment, due to the spin of unpaired electrons, and independent of the ambient magnetic field: they may impart ferro-magnetism to an otherwise non-magnetic material. 

For purposes of polarization studies, we are interested in steady-state solutions of eq. 6 to 8, i.e. those for which $\dot\omega_{i}=0$ ($i=1, 2, 3$). This is the case, in particular, if all 3 angular velocities are null.The system then reduces to $N_{i}=0$, or

\begin{eqnarray}
VH^2(\chi_{2}-\chi_{3})\mathrm{cos}q\mathrm{cos}r=0,\\
VH^2(\chi_{3}-\chi_{1})\mathrm{cos}r\mathrm{cos}p=0,\\
VH^2(\chi_{1}-\chi_{2})\mathrm{cos}p\mathrm{cos}q=0.
\end{eqnarray}

If all 3 susceptibilities are different, as with dia- or paramagnetism, then this implies that 2 of the angles $p,q,r$ are equal to 90 deg. The third is null, its corresponding principal axis being parallel to the field; of the 3 possible configuration, the system will select the one with the lowest potential energy $-\vec M.\vec H$. More specifically, for diamagnetic grains, the axis with the major $\mid\chi\mid$ will settle perpendicular to the field and that with the minor $\mid\chi\mid$, parallel to the field.

Another steady state configuration has the particle rotating with a  (finite but) constant angular velocity and magnetization simultaneously parallel to the field. This, of course, is the spinning top analog and usual case considered in the literature. For dia- or paramagnetic particles, it can only occur if the structure is also structurally as well as geometrically symmetric about the field; \bf otherwise, $s$ would not be null and Faraday braking is bound to occur and damp the rotation.\rm

In the case of ferro-magnetism, on the other hand, there are no finite susceptibilities, only a constant magnetic moment, say $M_{1}$. Then, $N_{i}=0$ translates into 
\begin{eqnarray}
M_{1}H\mathrm{cos}r=0,\\
M_{1}H\mathrm{cos}q=0,
\end{eqnarray}

which implies that $q=r=90$ deg so $p=0$, i.e. the permanent magnetic moment is parallel to the field, as expected.

The fact that steady state solutions exist does not imply that they can always be reached, whatever the combination of physical parameters and initial conditions. A major obstacle along the way is the exchange of rotation energy between the grain axes due to the gyroscopic torques, which are proportional to the differences between the 3 moments of inertia and hence increases with the asphericity of the grain. Another obstacle is the exchange of energy with the magnetic field, fueled by the magnetic torques, $N_{i}$, before they are nulled by the braking torques. These are, respectively, of the forms $\chi_{i}VH_ {i}H_{j}$ and $s_{k}\omega_{k}H$, overlooking sine factors in both. If $s$ and $\omega$ are not large enough to absorb the energy gained from the field during part of each cycle, then steady state cannot be reached. If, alternatively, the ratio

 $R_{1}=\frac{\chi_{i}VH_ {i}H_{j}} {s_{k}\omega_{k}H_{k}}$

 is small enough, the braking torque may overcome the magnetic torque for a while. However, when the rotation energy has fallen down to the level of the magnetic energy, the rotation motion is bound to turn into a pendulum motion. At that point, the maximum angular velocity is linked to the magnetic field strength by a relation of the form $I\omega^{2}= \chi VH^{2}$, so the ratio of the conservative term to the braking term becomes

 $R_{2}=\frac{(\chi VI)^{1/2}}{s}$. 

In the case of ferro-magnetic particles, this ratio is $(MI/H)^{1/2}/s$. \it Only if these ratios are smaller than about 10-100 can rotation and rocking be completely damped out and a static final state be reached.\rm The physics is made still more familiar if one notes that $R_{2}$ is, in order of magnitude, the inverse of the ratio of the damping to the critical damping coefficients for a damped oscillator. If the latter is too low, the grain will overshoot beyond its steady state position and execute oscillations about it: the lower the ratio, the longer these will endure. As the grain's  radius $a$ increases, its volume increases like $a^{3}$ and its moment of inertia like $a^{5}$. So, to maintain the ratios $R$ at the same value, $\chi$ must be reduced in the same ratio as $V^{-2/3}$.

Note the great disparity in magnitude between magnetic moments of different types: a single unpaired electron carries a magnetic moment of 1 Bohr magneton, i.e. $0.9\,10^{-20}$ erg/G, while a whole coronene molecule in a field of $3\,10^{-6}$ G carries a diamagnetic moment of $7\,10^{-35}$ erg/G.

Simple, calculable test particles will now be used to illustrate these considerations. This will help understanding the role and weight of parameters and initial conditions in determining alignment.

\subsection{Diamagnetic molecules}

\begin{figure}
\resizebox{\hsize}{!}{\includegraphics{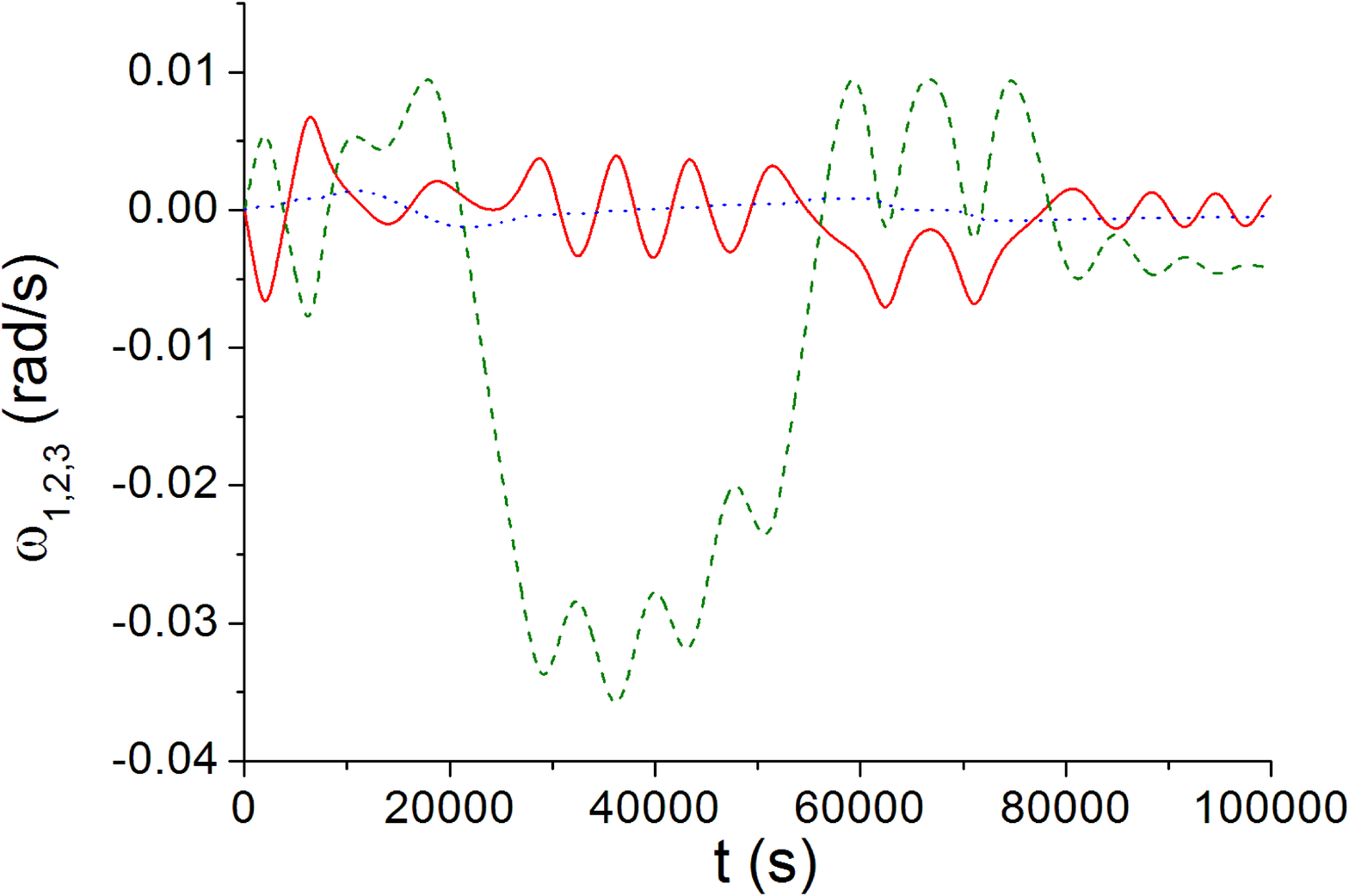}}
\caption[]{The mathematical model (eq. 5 to 8) describes the evolution in time of the 3 components of the angular velocities of neutral coronene (in the molecular frame, not the lab frame; full red, dashed olive and dotted blue for axes 1, 2 and 3), in a magnetic field of $3\,10^{-6}$ G, initially inclined with respect to the plane of the molecule. All 3 initial values are set to the same low value: $10^{-6}$ rad/ps.}
\label{Fig:w123coron}
\end{figure}

As a test dust particle, consider, for instance, coronene (24 C, 12 H), a diamagnetic molecule, and assume first that it is free of external interference except for the magnetic field. Geometry optimization using the semi-empirical PM3 code delivers the inertia moments: $I_{1,2,3}=2.57, 2.57, 5.14\, 10^{-37}$ g.cm$^{2}$, as well as the partial electric charge distribution, which gives $\Sigma(qx^{2})=\Sigma(qy^{2})=6.46$ e.$\AA{\ }^{2}$, and $\Sigma(qz^{2})=0$; hence, using eq. 4, we deduce $s_{1,2,3}=10.34, 10.34, 20.7\,10^{-36}$ escgs. Assume the molar susceptibilities to be the same as those  measured for the trioS molecule (Papoular 2016): $V\chi_{1,2,3}=-2.3, -2.39, -2 \,10^{-29}$ cm$^{3}$ (Gauss system), which are close to those of graphite. Finally, set the molecule in a magnetic field $3\,10^{-6}$ G, inclined to its plane and give it small, equal initial rotation speeds, $10^{-6}$ rad/s, about the 3 principal axes.

\begin{figure}
\resizebox{\hsize}{!}{\includegraphics{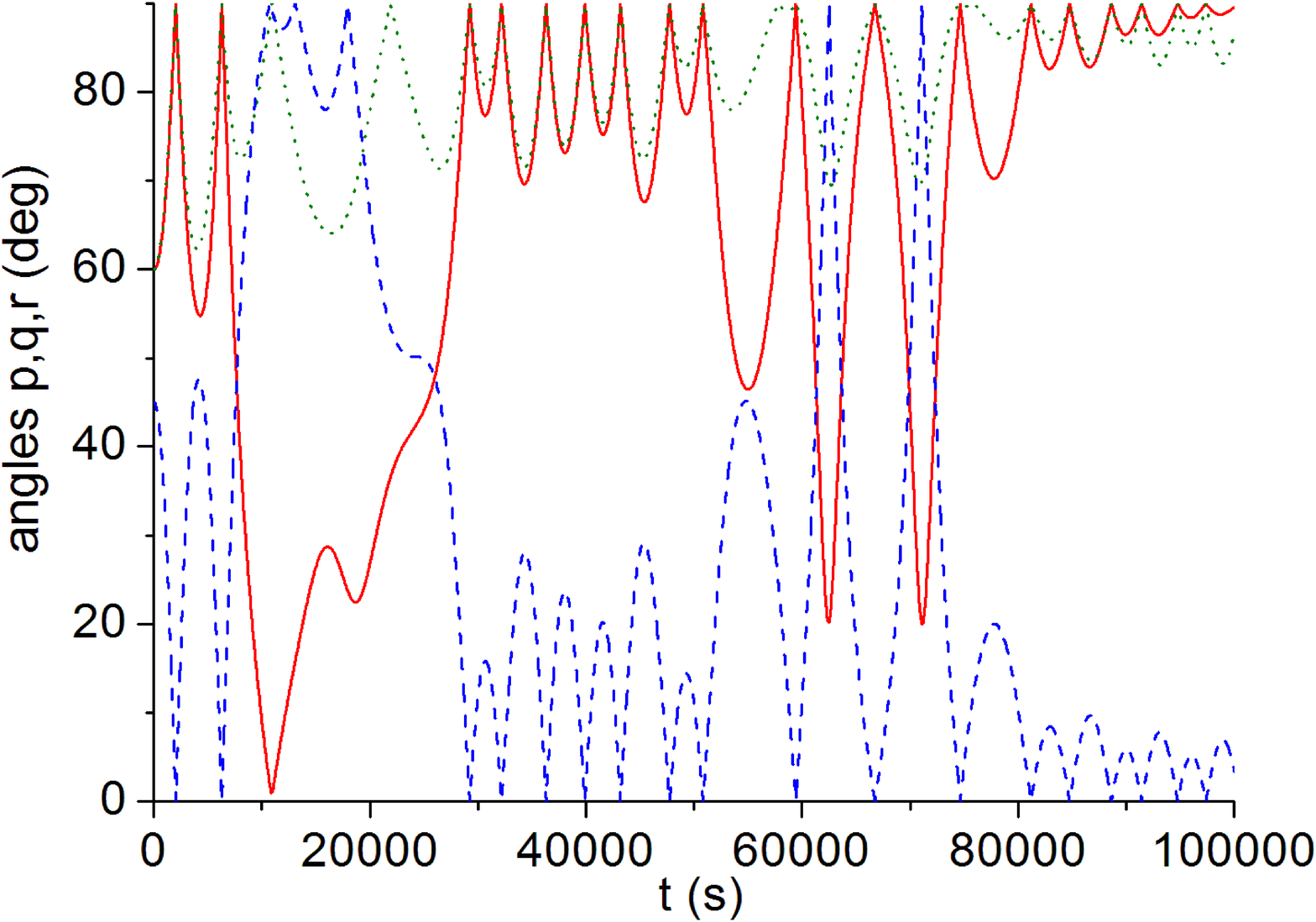}}
\caption[]{The figure displays the evolution in time of the angle made by each of the 3 principal axes of the dia-magnetic coronene with the magnetic field, $H=3\,10^{-6}$ G. As the rotation slows, axes 1 (solid red) and 2 (dotted olive) tend to be normal to the field while axis 3 (dashed blue) tends to become parallel to it, as predicted in Sec. 3.1.}
\label{Fig:pqrcoron}
\end{figure}

\begin{figure}
\resizebox{\hsize}{!}{\includegraphics{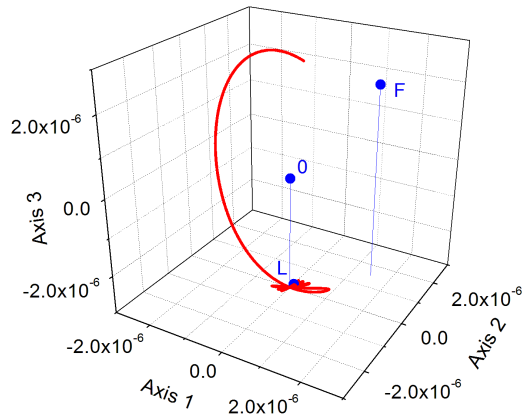}}
\caption[]{The figure displays the trajectory of the tip of the magnetic field in the internal frame of reference of the molecule (principal axes 1,2,3) at the end of the run, from $t=6\,10^{4}$ to $10^{5}$ (red). Also represented are the first (F) and last (L) points of the run (blue). $H=3\,10^{-6}$ G. Under the magnetic torques, the molecule changes its orientation relative to the field and ends up in a pendulum motion about the stable, static, ultimate configuration, where axis 3 (normal to the molecular plane) is parallel to the field. This demonstrates the achievability of alignment of a relatively small molecule, provided a balance obtains between magnetic susceptibility and the other parameters, as discussed in Sec. 3.1.}
\label{Fig:Hcoron}
\end{figure}

The numerical solution of system 6 to 8 delivers the evolution in time of the 3 principal rotation speeds, $\omega_{1,2,3}$, displayed in Fig. \ref{Fig:w123coron}. Figure
\ref{Fig:pqrcoron} illustrates the spatial motion of the molecule about its center of mass, through the  angles made by each principal axis with the field. Perhaps a more telling way of representing this evolution is by drawing the trajectory of the tip of the magnetic field in the internal frame of reference of the molecule (principal axes 1,2,3)
For the sake of clarity, this is done in Fig. \ref{Fig:Hcoron} only for a late part of the run, from $t=6\,10^{3}$ to $10^{4}$ s. In the end, although the motion is not completely quenched, the particle is indeed aligned relative to the field: here, with its plane nearly normal to the field.

Note that, in the diamagnetic case, the magnetic energy is $\mid\mu\mid H^{2}$, and so is always positive. The most stable orientation has the principal axis with the weakest susceptibility parallel or anti- parallel to the field. For coronene, this is axis 3, as in the figures.

\subsection{Ferro-magnetic molecules}
\begin{figure}
\resizebox{\hsize}{!}{\includegraphics{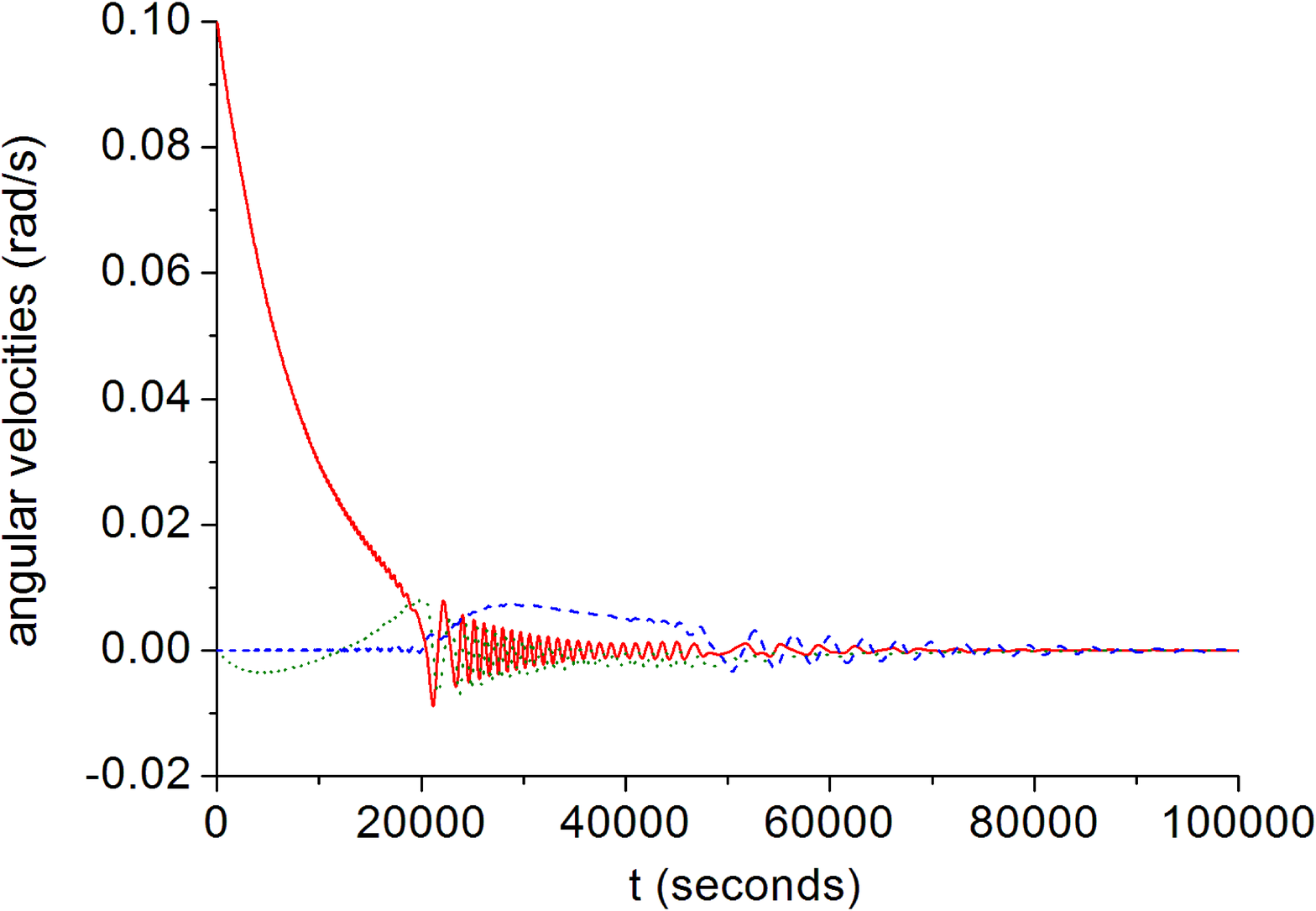}}
\caption[]{Evolution in time of the 3 components of the angular velocities of neutral ferromagnetic coronene (in the molecular frame) in a magnetic field of $3\,10^{-6}$ G, with a spontaneous magnetic moment $7\,10^{-37}$ erg/G along its axis 2: $\omega_{1}$ (solid red), $\omega_{2}$ (dotted olive); $\omega_{3}$ (dashed blue). The initial angular velocity along axis 1 is seen to be uniformly damped with a time constant 7500 s.}
\label{Fig:w123coronRa}
\end{figure}

\begin{figure}
\resizebox{\hsize}{!}{\includegraphics{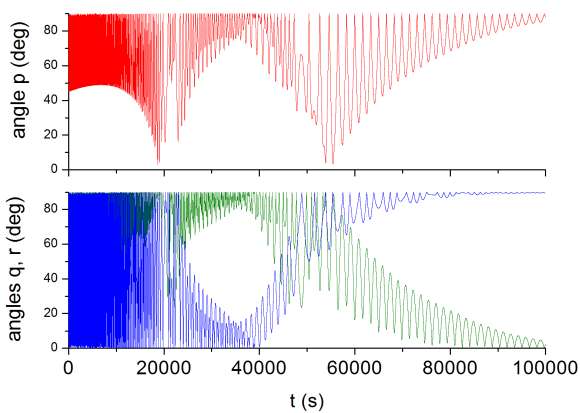}}
\caption[]{The figure displays the evolution in time of the angle made by each of the 3 principal axes of the ferromagnetic coronene with the magnetic field, $H=3\,10^{-6}$ G. Upper frame (red): angle $p$ of axis 1 with the field. Lower frame: angle $q$ of axis 2 (olive,full line), angle $r$ of axis 3 (blue dots). Under the gyroscopic and magnetic torques, the molecule rotates about the magnetic field in a complex way but, after rotation damping, ends up with axis 2 (bearing the magnetic moment) being very nearly parallel to the field.}
\label{Fig:pqrcoronRa}
\end{figure}

While the behavior of paramagnetic particles in a magnetic field does not significantly differ from diamagnetic ones \it mutatis mutandis \rm, the alignment of ferromagnetic particles relative to the field is distinct (Sec. 3.1 eq. 12-13). To illustrate this point, we take the same coronene molecule and assume, for the sake of the argument, that it has no diamagnetism and is instead given some ferro-magnetism by substituting carbon atoms with spin-bearing impurities. It has not proved possible to obtain alignment with a spontaneous magnetic moment of 1 Bohr magneton. Alignment becomes possible upon reducing this by a factor 100, to $\sim7\,10^{-37}$.This brought down the ratio of magnetic to braking torques down to $\sim$15. As expected, in the final steady state, reached after about $10^{5}$ s (Fig. \ref{Fig:w123coronRa}), the assumed spontaneous moment (along axis 2) stays parallel to the ambient field (Fig. \ref{Fig:pqrcoronRa}).

\subsection{Grains}

By grain is meant here a macroscopically homogeneous particle assumed to have grown by successive agglomerations of smaller particles of the same composition. The resulting structure is most likely to be disordered, albeit not necessarily perfectly disordered. Likewise, the shape cannot be perfectly spherical but, rather, spheroidal, prolate or oblate to some extent. 
 
 \begin{figure}
\resizebox{\hsize}{!}{\includegraphics{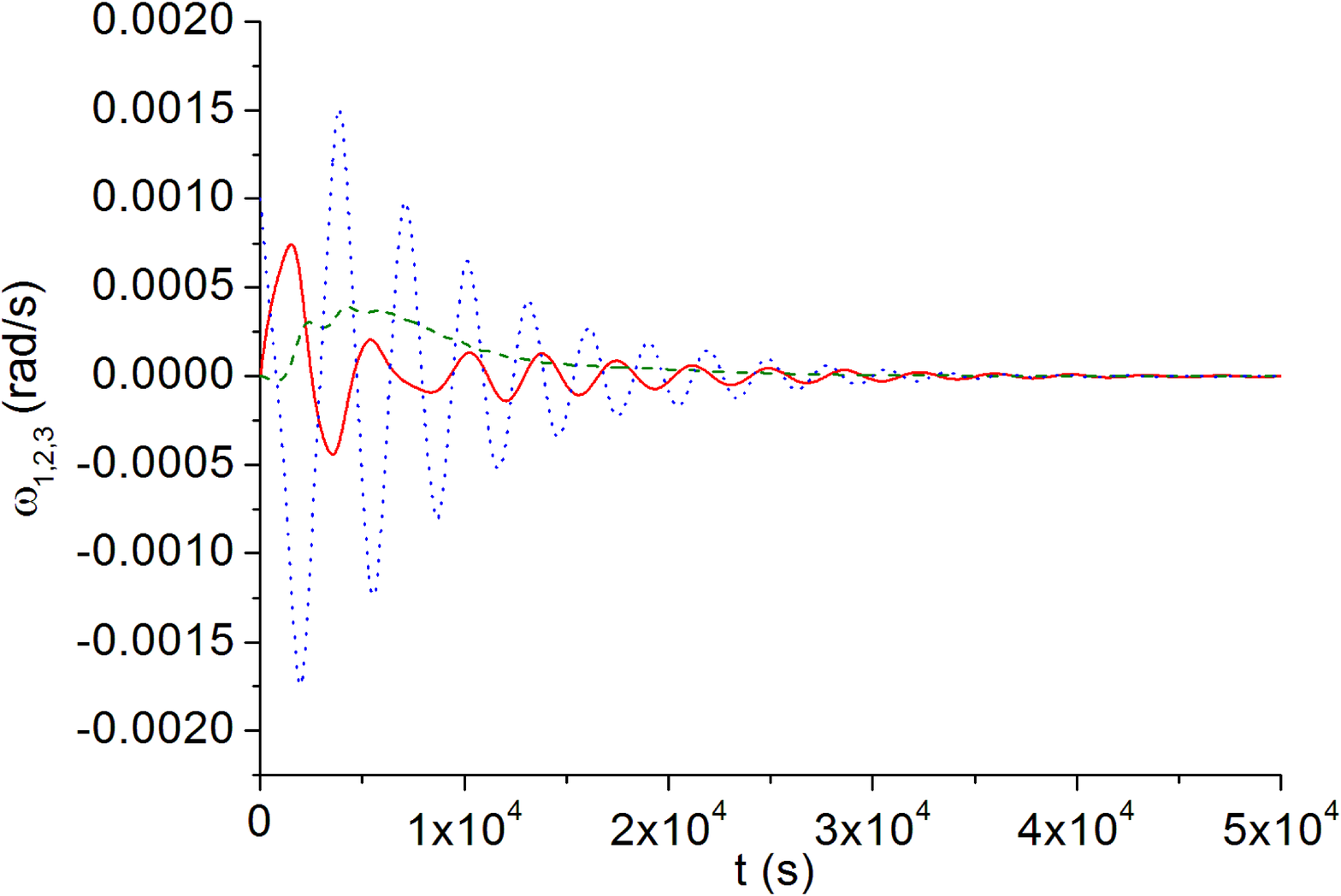}}
\caption[]{Evolution with time, and in the grain's frame, of the angular velocities (1: solid red, 2: dashed olive, 3: dotted blue)  of a ``big'' carbon-rich, paramagnetic, and slightly prolate grain  in a magnetic field of $10^{-5}$ G, Initial angular velocities along axes 1, 2, 3: $10^{-6}\,, 10^{-6}\,,10^{-3}$ rad/s respectively. After a moderate exchange of rotation energy between axes, the angular velocities decline with a time constant 8000 s.}
\label{Fig:w123grain}
\end{figure}

\begin{figure}
\resizebox{\hsize}{!}{\includegraphics{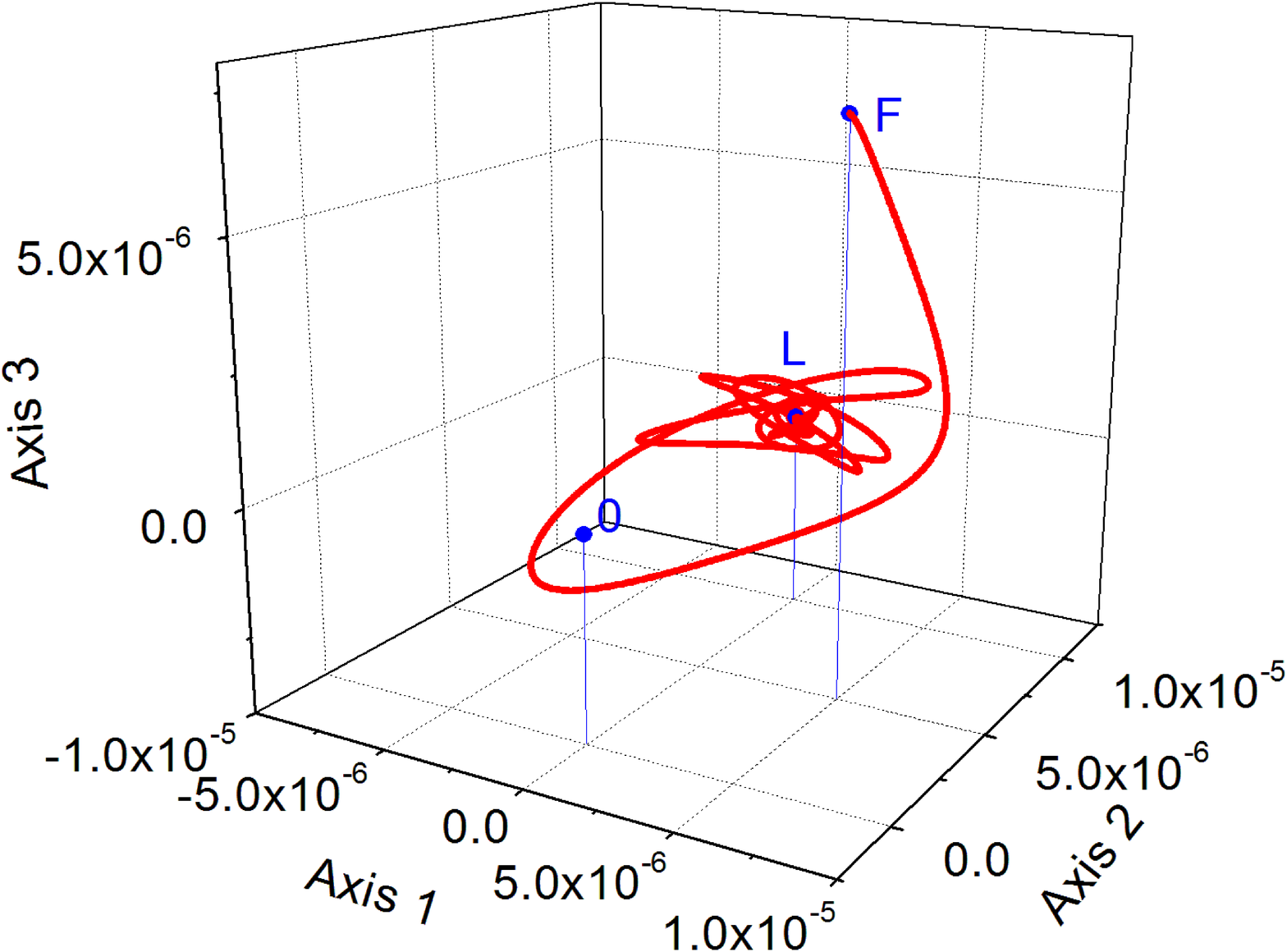}}
\caption[]{The figure displays the evolution in time of the tip of the magnetic field in the grain's frame. $H=10^{-5}$ G. Initial angular velocities along axes 1, 2, 3: $10^{-6}\,,10^{-6}\,,10^{-3}$ rad/s respectively. Also represented are the first (F) and last (L) points of the run (blue). O is the origin of coordinates. In the final configuration, at $t\sim10^{5}$ s, the grain's axis 2 is parallel to $H$.}
\label{Fig:Hgrain}
\end{figure}

\begin{figure}
\resizebox{\hsize}{!}{\includegraphics{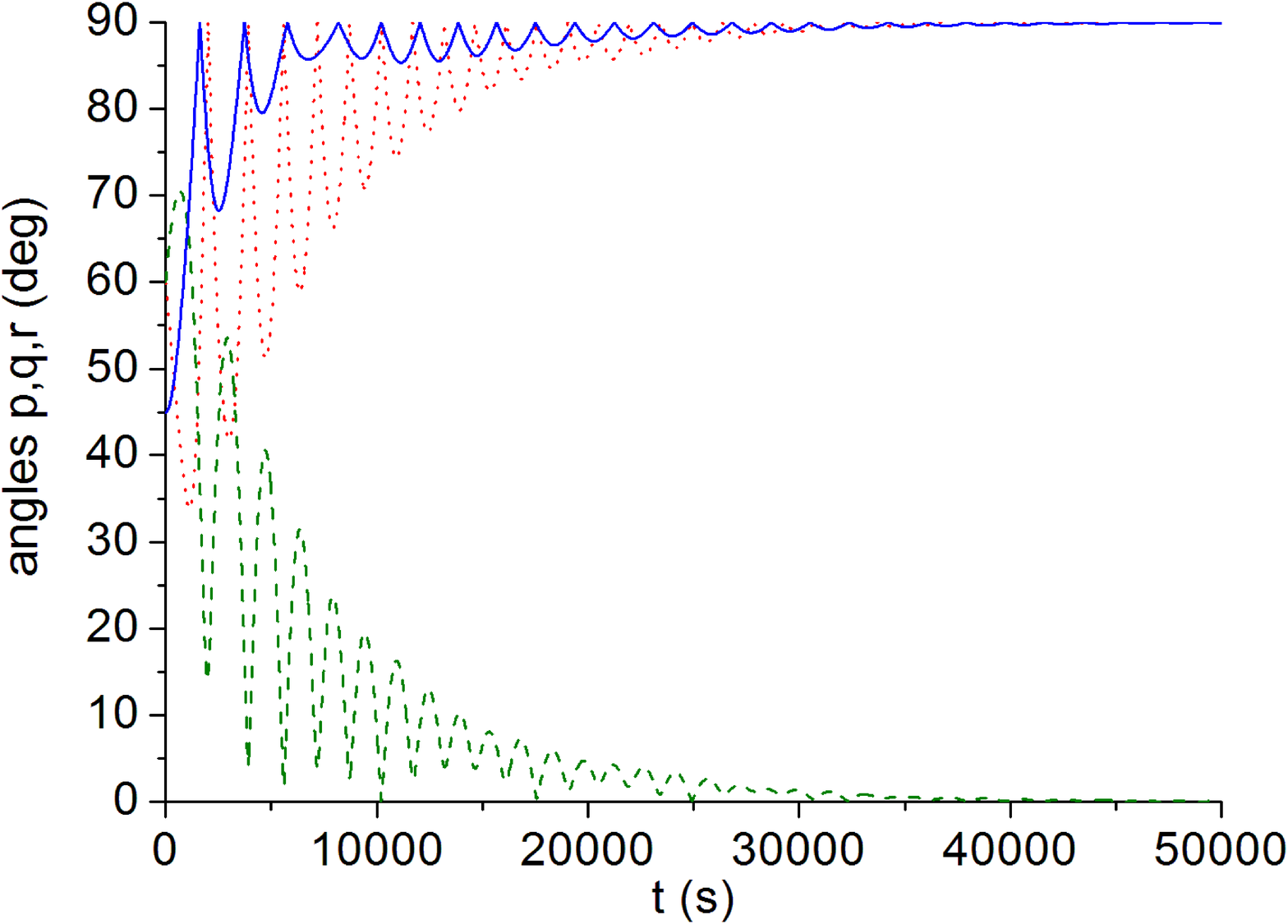}}
\caption[]{ The evolution in time of the angle made by each of the grain's 3 principal axes with the magnetic field: angle $p$ of axis 1 with the field (dotted red), angle $q$ of axis 2 (dashed,olive), angle $r$ of axis 3 (full blue). At the end of the run, the grain's axis 2, bearing the paramagnetic moment, is parallel to $H$.}
\label{Fig:pqrgrain}
\end{figure}

  As an example, take a prolate spheroidal grain, with linear dimensions about 30 times those of coronene, and moments of inertia 8, 8 and $5\,10^{-30}$ g.cm$^{2}$.
 The structural disorder considerably reduces the magnetic anisotropy, and hence the effective magnetic torques (and the ratios $R_{1}$ and $R_{2}$), to the advantage of the braking torques: every elementary induced magnetic moment is compensated by another elementary moment symmetric with respect to the field. 

\begin{figure}
\resizebox{\hsize}{!}{\includegraphics{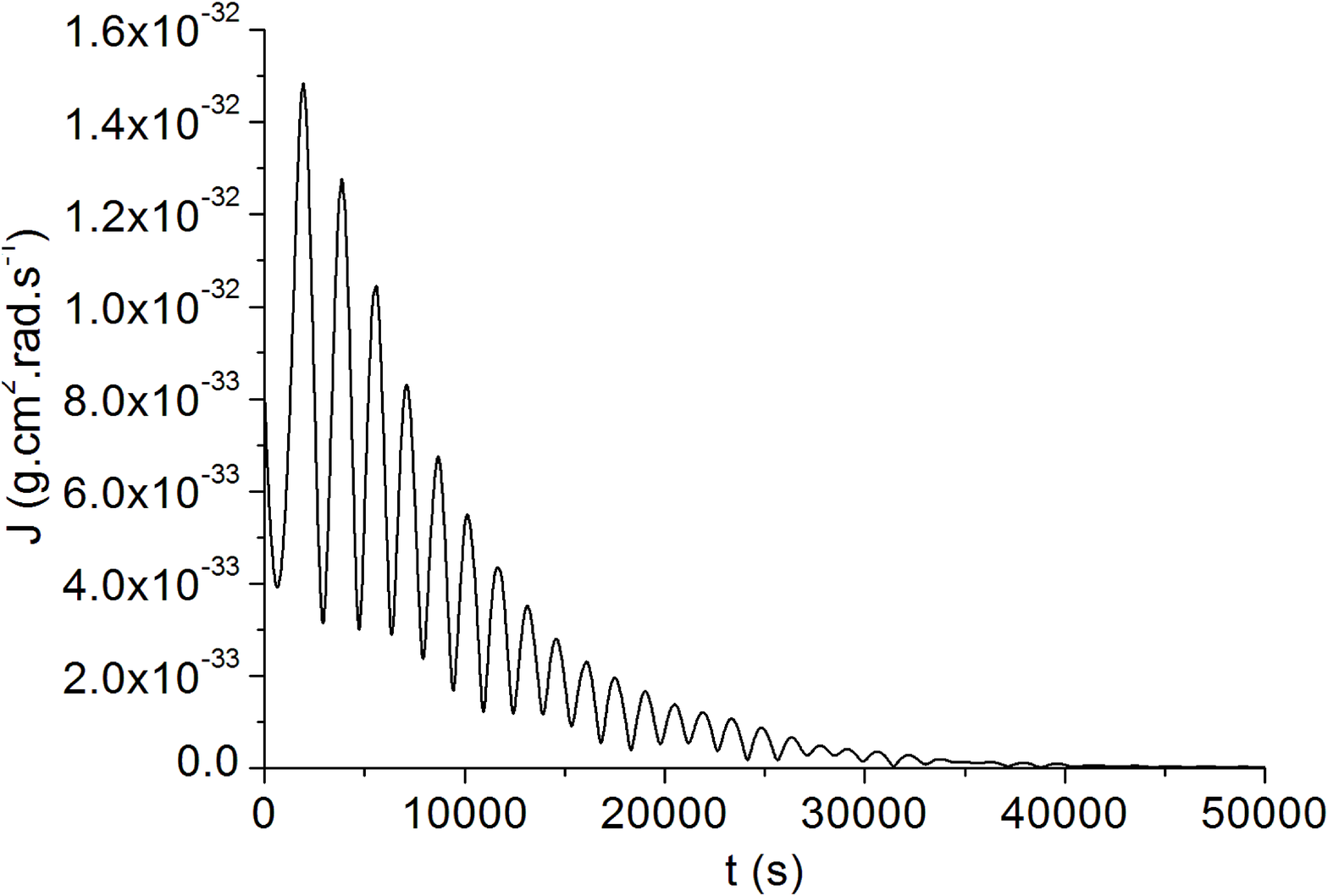}}
\caption[]{Variation with time of the magnitude of the angular momentum of the grain of Fig. \ref{Fig:w123grain} to 11.}
\label{Fig:Jgrain}
\end{figure}

For lack of magnetic data, assume, arbitrarily but guided by Sec. 3.1, that the grain is given, somehow, a dominant paramagnetic susceptibility along its short axis 2: $\chi V=2.3\,10^{-25}$ cm$^{3}$; also, for a convenient graphical presentation, assume a relatively high common damping factor $s=2\,10^{-28}$ for all 3 principal directions, to give a damping time $\tau_{b}=8000$ s. Also give the grain the arbitrary initial angular velocities $10^{-6}, 10^{-6}$ and $10^{-3}$ rad/s along axes 1, 2, 3 respectively and launch it in a field $10^{-5}$ G. The behavior of the grain governed by eq. 5-8 is illustrated in Fig. \ref{Fig:w123grain} to \ref{Fig:Jgrain}.

 Note how smoother it is, than that of molecules, because of the smaller disparity between the moments of inertia, which reduces rotation energy exchanges between the axes due to the gyroscopic torques (second terms in the left hand sides of eq. 5-7): after a limited exchange between axes 1 and 2, all 3 angular speeds steadily decrease, with the last one decreasing more quickly. 

These computations also show that the angular momentum is never constant: as it steers towards its ultimate direction, its average magnitude decreases all along (Fig. \ref{Fig:Jgrain}).

\begin{figure}
\resizebox{\hsize}{!}{\includegraphics{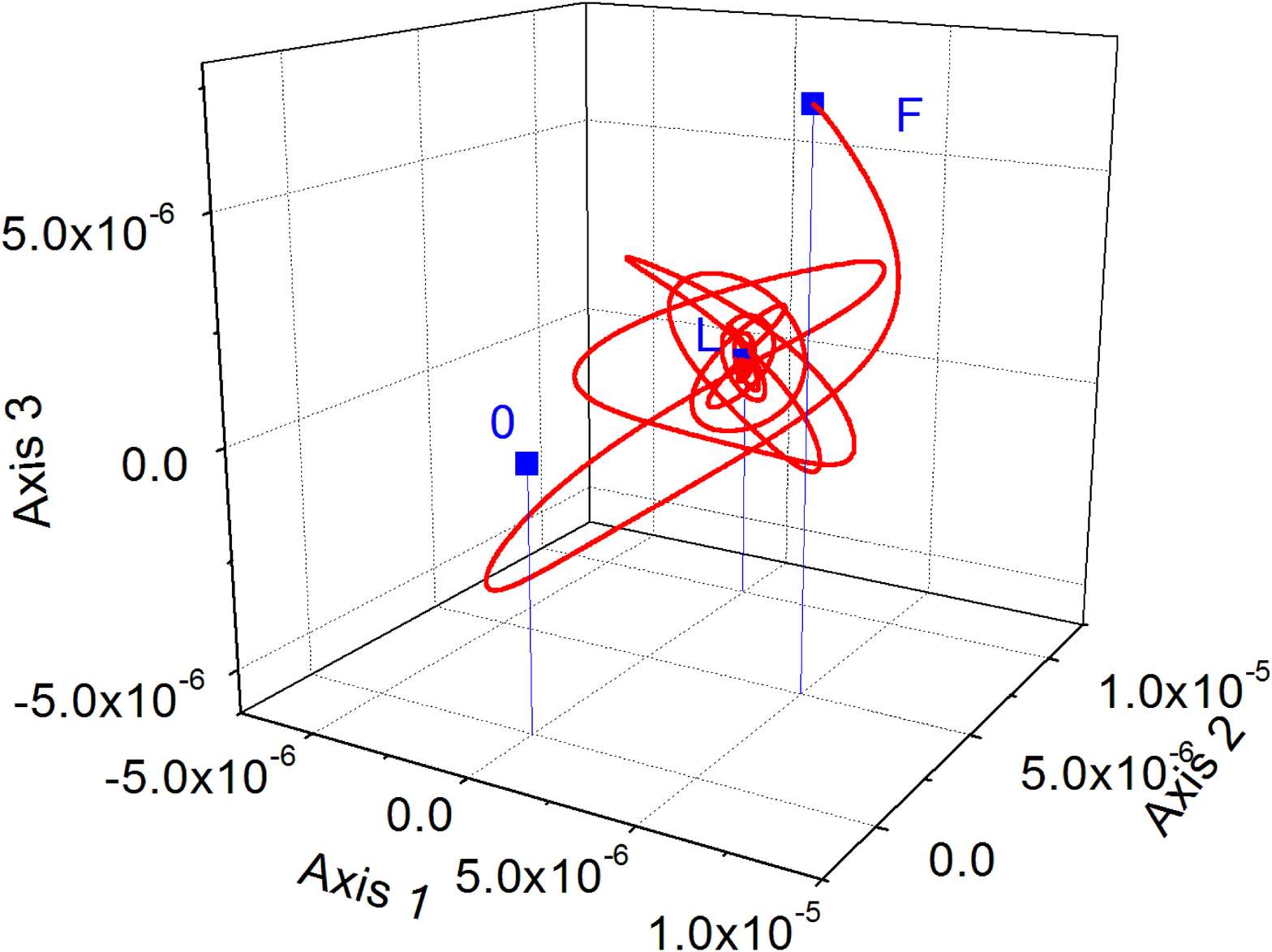}}
\caption[]{Same as Fig. \ref{Fig:Hgrain}, except that the magnetic moment along axis 2 is now constant ($2.3\,10^{-30}$, cgs units), as in ferromagnetic materials. Upon alignment, it ends up, as expected, parallel to the field. F,L: first and last points of the run.}
\label{Fig:Hferrograin}
\end{figure}

As a last example, take the same grain and dope it with iron so as to give it a spontaneous and constant magnetic moment along the same axis 2. Then, as predicted in Sec. 3.1, its evolution in the magnetic field ends up with this magnetic moment again parallel to the field (Fig. \ref{Fig:Hferrograin}), and the grain's long axis perpendicular to the field.
 
\bf Some anisotropy is obviously necessary for grain orientation. If the structural isotropy of the constituent material is nearly perfect, the remaining cause of anisotropy is the grain shape and this may determine the final orientation of the grain in the field. This is because of the demagnetizing field (see Osborne 1945), as shown next. Let $\chi$ be the (now) isotropic susceptibility, and $\vec H_{0}$, the external field. In response to the latter, an opposing field is created, of the form $-N\chi\vec H$, where $\vec H$ is the unknown internal field, and $N$ is a constant called the demagnetizing factor, and is determined, for each direction, by the ratios of the grain axes. The internal field is then found to be
\begin{equation}
\vec H=\frac{\vec H_{0}}{1+N\chi}.
\end{equation}
The computation of $N$ is excessively involved except for an ellipsoidal grain geometry. In this case, $N$ as a function of direction, can be described by an ellipsoid whose principal axes coincide with those of the grain. The sum of all three principal values of $N$ is  $4\pi$; $N$ is maximum along the shortest dimension and minimum along the longest. We seek now to define an apparent or effective susceptibility, $\chi_{eff}$. The induction $\vec B$ in the grain may be written either as $\mu_{eff}\vec H_{0}$ or as $\mu\vec H$, where $\mu=1+4\pi\chi$ is the permeability of the material, and $\mu_{eff}=1+4\pi\chi_{eff}$. Using the above relation between the fields, one readily finds
\begin{equation}
\chi_{eff}=(1-N/4\pi)\chi ,
\end{equation}
which indicates that the magnitude of the susceptibility is reduced by a factor between 0 and 1/3. Finally, from the definition of the magnetic potential energy and, hence, of the force or torque acting on the sample (see Van Vleck 1932), one concludes that the sample tends to move so as to maximize the volume integral of $\chi H^{2}$, in agreement with the behavior of the grains in the examples above (where $H_{0}$ is constant). If the grain material is isotropically diamagnetic, the grain will settle in the position where $N$ is maximum, i.e. with its long axis normal to the field, as required for polarization. \rm

\subsection{Further constraints on alignment }

Up to this point, the dust grain was assumed to be free of interactions with its environment. In fact, it is (mainly) subject to collisions with ambient H atoms. These may be of several types (see, for instance, Papoular 2005):

 a) Simple elastic recoil: in this case, the grain mass being much larger than that of the atom, the collision leaves only a very small fraction of its energy in the grain. Even this is quickly partitioned between all degrees of liberty of the latter. Among these, only 3 are assigned to rotation; vibrations absorb most of the energy (they are responsible for the specific heat) which is quickly radiated away in the infrared. As a consequence, the effect of an elastic atomic collision on rotation is negligible and thermodynamic equilibrium can hardly be completed.
 
 b) Chemical adsorption on a surface dangling bond of the grain. In this case, the kinetic momentum and energy of the atom are left in the grain. Again, because of the mass difference, the perturbation is assumed to be negligible.

c) Expulsion of a surface H atom. This requires improbably high ambient gas temperatures.

 d) Abstraction of an H atom bound to the grain, to form a hydrogen molecule, H$_{\mathrm{2}}$. Analysis of this case shows that most of the energy liberated by molecule formation (about 4.5 eV  or 110 kcal/mol) goes into the kinetic and vibrational energies of the free molecule, leaving less than 1 kcal/mol in the mother grain itself. While this is a small amount, the momentum given to the grain in reaction (rocket effect) suffices to send small molecules into relatively fast rotation. The interval between two successive rocket events is taken here as the time during which the particle can move freely.
 
 The probability of each of these occurrences depends, of course on the hydrogen coverage of the grain and the ambient atomic velocity. Assume these to be, respectively, 0.5 and $10^{5}$ cm.s$^{-1}$. Then, the probability of H$_{\mathrm{2}}$ formation upon a direct hit on an H surface atom (case (d)) is estimated at about 0.5, so the effective cross-section for this reaction is about 1/4 of the geometric cross-section.
 
 Finally assume the ambient H density to be 20 cm$^{-3}$ and the grain geometric cross-section,100 $\AA{\ }^{2}$, then, the average interval between rocketing impacts is $t_{r}=2\,10^{8}$ s. \it Whether rotation is ultimately damped or the grain has settled into a steady state, with the angular velocity vector making a definite, constant angle with the magnetic field, the whole process must be terminated in a time much shorter than the interval $t_{r}$.\rm The alignment condition is therefore met in the particular case of coronene in Sec. 3.2 and the big disordered grain in Sec. 3.4.

The alignment condition just discussed implies, of course, the condition $\tau_{b}<<t_{r}$.

\bf It is also necessary to inquire whether, torques which were not included above in the dynamics equations could, if active, interfere with the alignment process. While this can be answered in particular cases by implementing their mathematical expressions in Eq. 5-8, the qualitative effects can be predicted beforehand.

 Any other purely braking torque, included in Eq. 5-8, would only reduce $\tau_{b}$ and favor alignment between successive rocketing events.

On the other hand, conservative torques, like the $N$'s above, are more problematic as they may increase the ratios $R$ (Sec. 3.1) beyond their critical values and, possibly, bar access to steady states. Note that the  $R$ constraint is not specific to the present model and need be taken into account with braking torques other than Faraday's. This is not the case for the purely mechanical alignment in gas flows (Gold 1952): its contribution is likely to reduce $\tau_{b}$ provided the direction of the flow is adequately correlated with that of the magnetic field.

The interference of other torques may become still more worrying if they accelerate grain rotation instead of braking it, which is the case, for instance, of radiative torques (RATs). The situation can be illustrated by reference to the generic work of Draine and Weingartner \cite{dw}. These authors show that, in the presence of anisotropic irradiation, the grain can be set in suprathermal rotation, provided it has an adequate shape. Three adequate grain shapes were proposed and detailed dynamics in a field were given for Model I and illustrated in their Fig. 6. This shows that, for a wide range of initial conditions, the grain is indeed accelerated to suprathermal angular velocities and settles in a steady state. Obviously, under strong anisotropic radiation, as in the neighborhood of a star, this process would overcome our alignment scenario, provided it could be completed in a short time. However, Fig. 6, just mentioned, also shows that the time requested, for the most favorable initial conditions, is already of order $10^{5}$ y, very much longer than the rocketing time constraint already imposed on our model  ($t_{r}\sim 10$ y).

Another set of orientation times for RATs had previously been computed by Dolginov and Mytrophanov  \cite{dol} for typical HI and HII regions, and at 1AU of the Sun. 
Only in this last case is this time shorter than  $t_{r}$ and as short as $10^{3}$ s.\rm

\section{Polarization of light by aligned grains}

\begin{figure}
\resizebox{\hsize}{!}{\includegraphics{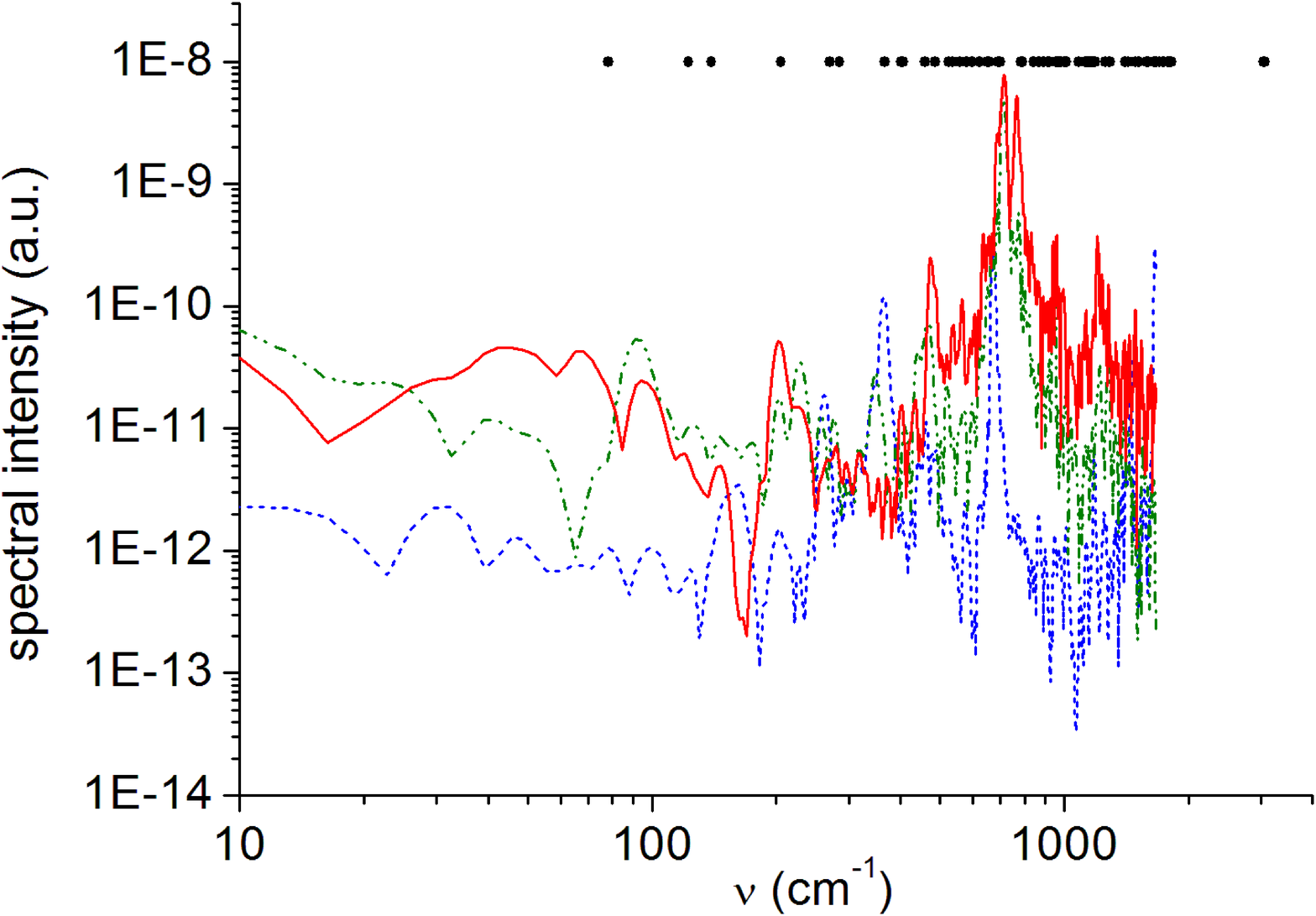}}
\caption[]{The spectral intensity of each and every IR coronene mode for electric field in each of the 3 principal directions of the molecule. \it Solid red, dash dot dot olive,  dashed blue\rm: axes 1, 2 and 3 respectively. The modes are preferentially polarized in the molecular plane.}
\label{Fig:coronirspec}
\end{figure}

The polarization $P$ of light emitted/absorbed by molecules and small grains is mainly intrinsic, i.e. determined by their structure. For large grains, made by random agglomeration of smaller ones, the structure is nearly isotropic, so the main polarization factor becomes their shape. As an example, consider the coronene molecule in its ultimate steady state (Sec. 3.2); in order to determine its intrinsic polarization, we return to the chemical code and give the coronene a sufficiently high temperature for a time long enough that all vibrations are excited. All information about these is registered in the resulting fluctuations of the molecular dipole moment; the evolution of its 3 components  is therefore  followed during about 10 ps, with an elementary time step of 0.03 ps. The amplitude Fourier transforms of the autocorrelation of these components, $rc_{1,2,3}$, are proportional to the square of the electric vectors of the corresponding emissions. These are plotted in Fig. \ref{Fig:coronirspec}; for comparison, the upper points at a common ordinate $10^{-8}$ indicate the wavenumber of the molecular normal modes as computed independently by the chemical code. The component along axis 3 is distinctly weaker than the other two, which shows that the atomic vibrations stay mainly in the molecular plane, indicating a clear intrinsic polarization. 

\begin{figure}
\resizebox{\hsize}{!}{\includegraphics{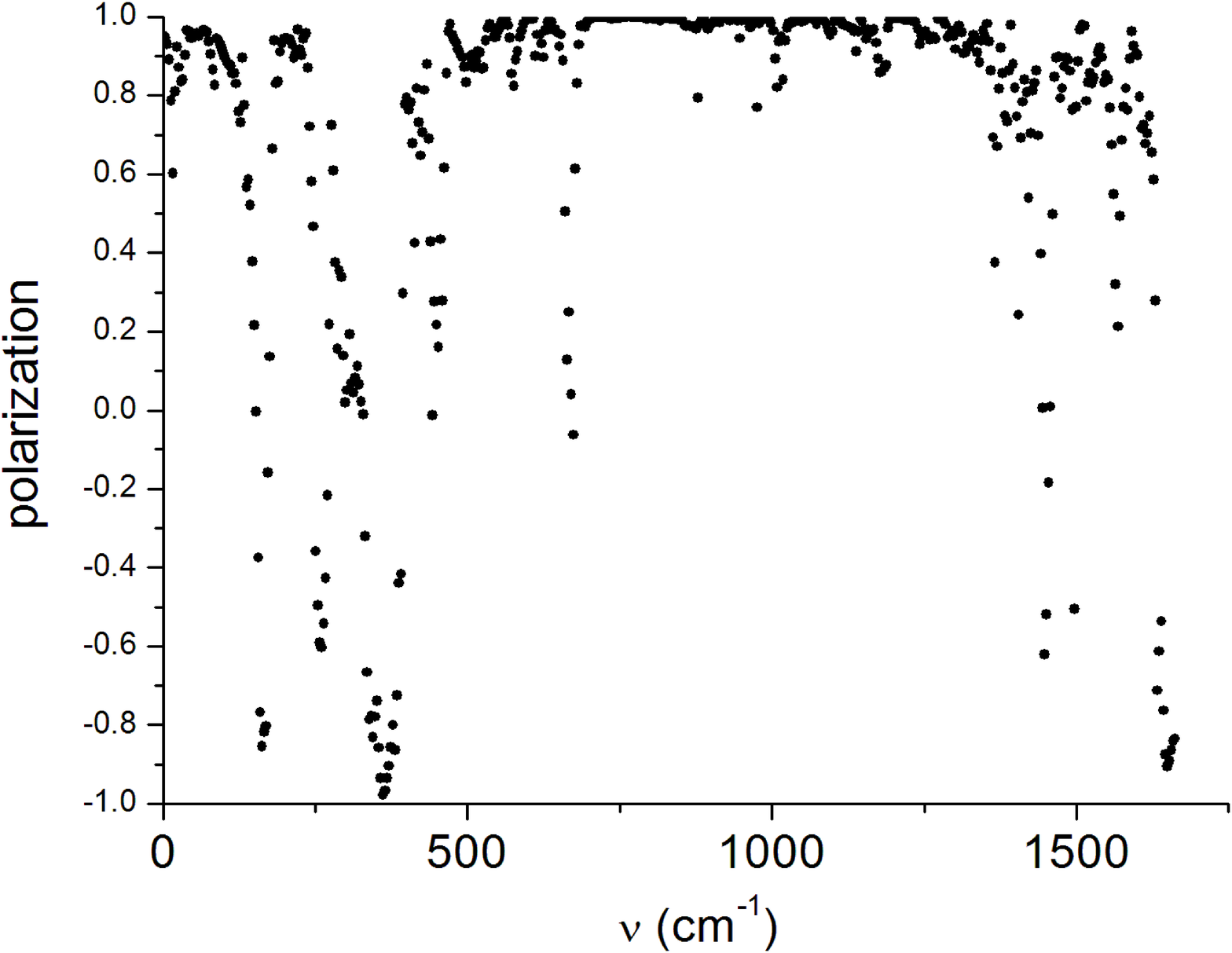}}
\caption[]{The \it intrinsic \rm polarization of far infrared emission from the diamagnetic coronene (Sec. 3.2) in its ultimate configuration, along a sight lign orthogonal to the magnetic field and electric field parallel to principal axis 2. Each resolved frequency is represented by a dot, and has a different polarization degree. The average over all modes is 0.71, with a standard deviation 0.5.}
\label{Fig:Pcoron}
\end{figure}

More precisely, for an observation sight line along axis 1, we compute

\begin{equation}
P_{i,\nu}=\frac{rc_{j,\nu}-rc_{k,\nu}}{rc_{j,\nu}+rc_{k,\nu}},
\end{equation}

and the corresponding expressions for sight lines along axes 2 and 3. Figure \ref{Fig:Pcoron} is a scatter plot of $P$, as a function of mode wavenumber, $\nu$, for the vibration electric field along axis 2. As expected, each resolved wavenumber is polarized to a different degree: This is characteristic of the mid- and far-infrared domains because of the multiplicity of distinct bands that they harbor. As an indication, the average polarization factors is about 0.71, and 0.51 and 0.41 for sight lines parallel to axes 1 and 3 respectively. In the case of Fig. \ref{Fig:Hcoron}, the plane of the coronene ultimately settles nearly orthogonally to the field (axis 3 parallel to $\vec H$). Then, the sight lines of interest are parallel to axis 1 and 2, respectively. In both cases, the intensity is higher perpendicular to the field (P$>$0).

\begin{figure}
\resizebox{\hsize}{!}{\includegraphics{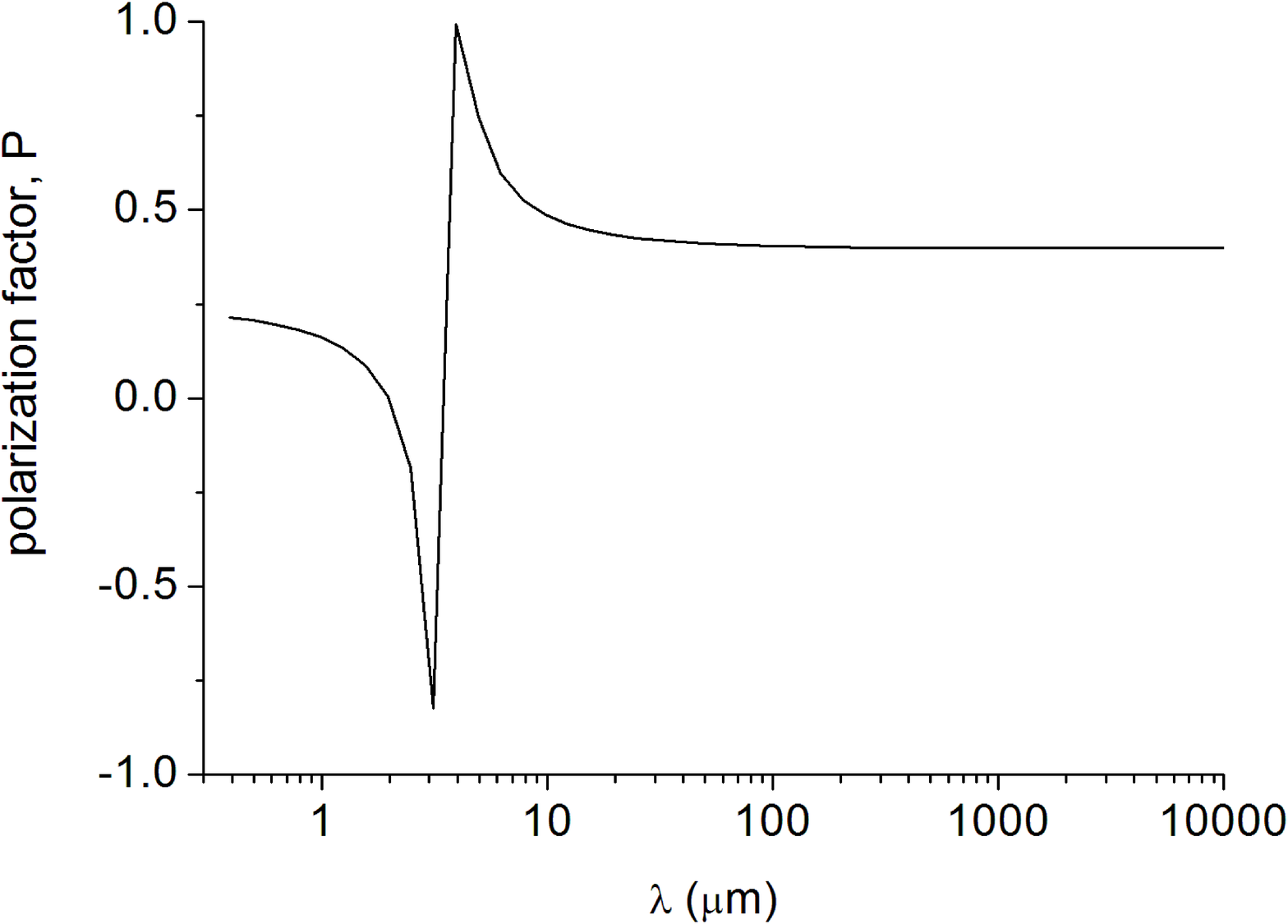}}
\caption[]{The polarization of light transmitted by the graphite grain (Sec. 3.4, ellipticity: 0.71) in its ultimate configuration (longer axis perpendicular to the field), along a sight line orthogonal to the magnetic field, with the  electric field parallel to the longer principal axis. The curve is only valid in the range where the Rayleigh/Gans approximation applies. This is the case for a grain size of 100 \AA{\ }.}
\label{Fig:Pgrain}
\end{figure}

Large dust particles cannot be treated by chemical modeling, so their optical properties must be determined by measurement. On the other hand, even grain sizes of order 1000 \AA{\ } are still smaller than the millimeter wavelength recently measured, for instance, by the COBE and Planck satellites. Their optical cross-sections are, therefore, a complicated mix of optical constants and geometric cross-sections, which must be treated by the Rayleigh-Gans theory (see, for instance, Davis and Geenstein 1951). More recently, this theory was presented by Bohren and Huffman \cite{bh} in a very convenient form which we apply here to the big grain given as an example in Sec. 3.4. In the choice of the constituent material, we take advantage of the recent measurements on graphene by A. Kuzmenko et al. \cite{ak}, which were then used by Papoular and Papoular \cite{pap14} to compute the optical properties of graphite from the UV to the far IR. The Appendix details how these data were used to deduce the polarization of light by our big grain. This is drawn in Fig. \ref{Fig:Pgrain} as

\begin{equation}
P=\frac{C_{\bot}-C_{\parallel}}{C_{\bot}+C_{\parallel}}\,,
\end{equation}

where $C$ designates the grain absorption cross-section as a function of wavelength, in its ultimate position relative to the field, for incident light polarized perpendicular ($\bot$) or parallel ($\parallel$) to $\vec H$, as computed in the Appendix. \bf This quantity is akin to the polarizing effectiveness defined by Spitzer \cite{spi78} and to the polarizing efficiency defined by Voshchinnikov \cite{vos}: polarization divided by total extinction at the same wavelength.\rm Here, $P$ is referred to the extinction caused only by the grain itself. It is about 0.2 in the UV/vis and 0.4 in the mid and far IR; this is only meant as an estimation of the contribution of graphite grains, ignoring other dust, such as silicates and silicon carbide. The discontinuity in between is a Fr$\ddot{o}$hlich , or surface, mode depending on the grain ellipticity (see Bohren and Huffman 1983), and linked with the transition from dominant bound $\pi$ electrons to dominant Drude free (conducting) electrons, a characteristic of graphite.  If, in the final, stationary, state, the grain's long axis is randomly distributed around the magnetic field, then $P$ should be halved.

To a factor of order 1, $P$ is also the polarization fraction $p$ defined by Ade et al. \cite{ade} in their analysis of the Planck measurements of the polarization of thermal dust emission in the Galactic plane. At a radiation frequency of 353 Ghz, this is found to lie between 0 and 0.2. The polarization was also found not to change significantly down to 100 GHz. Figure 16 above is quantitatively compatible with these observations.

\bf For large grains and short starlight wavelengths, the Rayleigh /Gans approximation may no longer apply. The full electromagnetic theory must then be used, with boundary conditions at the grain surface. This was done for spherical grains (Mie's theory) and for a few special grain shapes, e.g. long cylinders (Spitzer 1978, Bohren and Huffman 1983) and, most importantly for the present issue, for prolate and oblate spheroidal grains (see Asano and Yamamoto 1975). The variation of polarization with wavelength then depends, not only on the grain's shape and optical properties, but also on its size. For sizes increasing beyond the wavelength, the differences in parallel and perpendicular extinction efficiencies and, hence, the accompanying polarization roll down monotonously.\rm

\section{Conclusion}
The main results of this work are as follows.

1) An expression is given which allows one to compute the Faraday rotation braking torque and the damping time form first principles, for any given atomic structure.

2) It is then possible to write down explicitly the equation of motion of the structure in a magnetic field. Given initial conditions, this can be solved numerically, thus  avoiding any simplifying assumption. \bf Any other damping process may be included in the last terms of Euler's eq. 6-9, provided it can be expressed in quantitative form. This would only reduce the braking time. \rm

3) One can then study independently the effects and importance of the gyroscopic, conservative magnetic and braking torques, as well as of the initial conditions (shape, size, angular velocities, etc.) upon alignment. This leads to the following guide lines in the search for candidate polarizers.

4) Molecules and grains can align with the magnetic field provided the gyroscopic and magnetic torques are, in order of magnitude, smaller than the braking torque. Hence, low grain ellipticity and magnetic susceptibility are preferable. Both para- and dia-magnetism can do.

5) High initial angular velocity is not mandatory although it may help the braking beat the magnetic torque.

6) Four examples illustrate these statements: a diamagnetic and a ferromagnetic molecule, and a larger, homogeneous but disordered, graphite grain doped with a  para- or a ferromagnetic impurity.

7) In general, the angular moment is not constant during alignment. Because of the non- linearity of the equations of motion, numerical simulation is usually preferable to simplifying analytical assumptions.

8) Polarization cannot be treated in the same way for molecules and for grains  Molecules are naturally anisotropic and diamagnetic to some degree and can, therefore, align in the interstellar field. Molecular vibrations have intrinsic polarization, independent of their orientation, which must be taken into account for any given alignment. 

 \bf 9)  Being due to the rotation of electrons around their nuclei, dia-magnetism is the most common form of magnetism. It is usually dominant in the absence of paramagnetic ions, such as Fe, Cr, Mn, Ni, Ti. Astrophysically interesting diamagnetic materials are: carbon-rich structures (SiC, graphite, etc.) as well as non-iron bearing  silicates (like fosterite, which has the structure of olivine) and ice. Their susceptibilities are well documented in the literature. \bf Grains made of such material will align with their longer axis perpendicular to the field within a year or so, depending on their size.
 
 It is instructive to note that McMillan and Tapia \cite{mac} concluded that their study of polarization in the Cygnus OB2 association pointed to a weakly absorbing polarizing dust rather than to an iron-bearing dust like magnetite, which confirms other evidence to the same effect (Spitzer 1978).

10) Ferro-magnetism is associated with unpaired electron spins of certain atomic species which, in randomly distributed microscopic domains, happen to be  spontaneously aligned (see Kittel 1986,  Aharoni 2007, Chikazumi 2009). The minimum size of a domain is of order 100 \AA{\ }, which requires large grains with magnetic impurities concentrated in chunks. Moreover, it is not clear, at this time,  how the magnetization of such chunks can be set perpendicular to the grain long axis for the latter to settle perpendicular to the field.

11) The grain size distribution is limited at its lower end by the need of at least a few tens of normal vibration modes to absorb the initial rotation energy of the grain  (Sec. 2.3). This sets the minimum grain size at 10 to 20 \AA{\ }.  On the other hand, 
the braking time, $\tau_{b}$,  which scales like $I/s$ (Eq. 3), increases quickly with $a$. The maximum grain size is limited by the condition that $\tau_{b}$ be much shorter than the interval, $t_{r}$, between rocket-inducing collisions with ambient H atoms (Sec. 3.5). In between, the optimum size for polarization is in the order of 100 to 1000 \AA{\ }.

12) Different alignment models naturally rely on different assumptions or domains of environmental parameters. A comparison becomes possible whenever a model performance is quantified, for instance, by specifying the braking time and the time length of the whole process, as is done here, like in the earlier classical models (Davis-Greenstein, Spitzer, Purcell, Dolginov-Mytrophynov); see in particular Table 1 of the latter authors (1976). In this, only radiative torques near the sun yield shorter times than the present model. Using Table 1 of Andersson et al. \cite{and}, one can at best make a qualitative comparison by assigning ``pluses" or ``minuses" to performances under different criteria. There, again, the present model seems to fare honorably.\rm

\bf Acknowledgments. \rm I thank the anonymous reviewer for several comments which led to a significant improvement of the text.

\section{Appendix: grain cross-section for light polarization}

The treatment of polarization by spheroidal grains under the Rayleigh-Gans approximation ($a<\lambda$), as presented by Bohren and Huffman \cite{bh}, will be used here. In their Chapter 5, these authors give the absorption cross-section for a given alignment as

\begin{equation}
C_{abs}(\omega,L)=\frac{v\gamma\omega_{p}^{2}}{c\sqrt\epsilon_{m}}f(\xi,L)\frac{\omega^{2}}{(\omega^{2}-\omega_{s}^{2})^{2}+\gamma^{2}\omega^{2}}\,,
\end{equation}

where $v, \epsilon_{m}, \omega, \omega_{p}, \gamma$ are, respectively, the grain volume and the real dielectric constant of the ambient medium, the light frequency, the grain plasma frequency and the line width of its resonance frequency, $\omega_{t}$  (frequencies in eV). Also,

\begin{eqnarray}
f(\xi,L)=\frac{1}{(1+L(\xi-1))^{2}},\\
\omega_{s}^{2}=\omega_{t}^{2}+\frac{L\omega_{p}^{2}}{\epsilon_{m}+L(\epsilon_{0e}-\epsilon_{m})}\,,
\end{eqnarray}

where $\xi=\epsilon_{0e}/\epsilon_{m}$ and $L$, a depolarization factor defined by the grain ellipticity $e$ for alignment of the light's electric field along each of its 3 principal directions and obeying the relation $L_{a}+L_{b}+L_{c}=1$. For a prolate spheroid with semi-axes $a=b<c$,

\begin{equation}
e=(1-\frac{b^{2}}{a^{2}})^{1/2}\,.
\end{equation}

In the example of Sec. 3.4, the grain has $a=b=2.84\,10^{-6}$ and $c=4.1\,10^{-6}$ cm, so $v=1.33\,10^{-16}$ cm$^{3}$ and $e=0.71$. The theory then delivers

\begin{equation}
L_{a}=L_{b}=0.373,\,\,L_{c}=0.2\\55\,,
\end{equation}

which apply, respectively, to electric polarization along the short and long axes.

The grain is assumed to be made of randomly oriented chunks of graphite of the same size. The absorption of light by graphite is essentially the absorption of that component whose electric vector is normal to the basal plane. The dielectric functions for this component were shown to be correctly fitted by the sum of a Lorenz and a Drude dielectric functions, for the bound ($\pi$) and free electrons, respectively. The total absorption cross-section is the sum of the 2 corresponding cross-sections.  Table 1 gathers the 2 corresponding sets of quantities that enter the expressions above for each type of dielectric function and for each value of $L$. These are used to finally compute the total absorption cross-sections parallel and normal to the field $\vec H$: $C_{\parallel},\,C_{\bot}$ (Fig. \ref{Fig:Pgrain}).

\begin{table*}[ht]
\caption[]{Graphite parameters for absorption cross-section, eq. 16}
\begin{flushleft}
\begin{tabular}{llllllllllll}
\hline
L & $\epsilon_{0e}$ & $\epsilon_{0e\pi}$ & $\xi$ & $\xi_{\pi}$ & f & $\omega_{t}$ & $\omega_{p}$ & $\omega_{s}$ & $\omega_{t\pi}$ & $\omega_{p\pi}$ & $\omega_{s\pi}$\\
\hline
0.255 & 1 & 6 & 1 & 6 & 0.193 & 0 & 0.617 & 0.312 & 4.42 & 9 & 5.35\\
\hline
0.373 & 1 & 6 & 1 & 6 & 0.122 & 0 & 0.617 & 0.377 & 4.42 & 9 & 5.48\\
\hline
\end{tabular}
\end{flushleft}
\end{table*}


\begin{thebibliography}{}
    \bibitem[2015]{ade} Ade P. et al. 2015, Planck intermediate results XIX, A\&A 576, A104 and A107
    \bibitem[2007]{aha}Aharoni A. 2007, Introduction to the theory of ferromagnetism, 2nd edition, Oxford Univ. Press, Oxford
   \bibitem[2015]{and}Andersson B.-G., Lazarian A. and Vaillencourt J. 2015, ARAA, 53, 501
   \bibitem[1975]{asa}Asano S. and Yamamoto G. 1975, Appl. Optics 14, 29
 \bibitem[1983]{bh}Bohren C. and Huffman D. 1983, Absorption and scattering of light by small particles, John Wiley and Sons, New York
  \bibitem[2009]{chi}Chikazumi S. 2009, Physics of ferro-magnetism, Oxford Univ. Press, Oxford
\bibitem[1951]{dg}Davis L. and Greenstein J. 1951, ApJ. 114, 206
  \bibitem[1976]{dol}Dolginov A. and Mytrophanov J. 1976, Astroph. Sp. Sc.  43, 291
    \bibitem[1997]{dw}Draine B. and Weingartner J. 1997,  ApJ. 480, 633
  \bibitem[1998]{dra}Draine B. and Lazarian A. 1998, ApJ 508, 157
    \bibitem[1952]{gol}Gold T. 1952, MNRAS 112, 2015
 \bibitem[1980]{gol}Goldstein H. 1980, Classical Mechanics, AW Inc.
 \bibitem[1986]{kit}Kittel C. 1986, Introduction to solid state physics, J. Wiley $\&$ Sons, N. Y.
 \bibitem[1966]{kub}Kubo R. 1966, Rep. Prog. Phys. 29 Part I, 255
\bibitem[2008]{ak}Kuzmenko A., van Heumen E., Carbon F. and van der Marel D. 2008, Phys. Rev. Lett. 100, 117401
  \bibitem[2007]{laz}Lazarian A. and Hoang T.  MNRAS 2007, 378, 910
\bibitem[1977]{mac}McMillan R. and Tapia S. ApJ 1977, 212, 714
  \bibitem[1945]{osb}Osborn J. 1945, PR 65, 351
  \bibitem[2014]{pap14}Papoular R. J. and Papoular R. 2014, MNRAS
 \bibitem[2016]{pap16}Papoular R. 2016, MNRAS 457, 1626.
 \bibitem[2017]{pap17}Papoular R. 2017, MNRAS, doi:10.1093/mnras/stx100; (P17)
 \bibitem[1979]{pur79}Purcell E. 1979, ApJ. 231, 404 
 \bibitem[1984]{rei}Reif F. 1984, Fundamentals of statistical and thermal physics, Mc Graw-Hill Inc., Singapore
  \bibitem[1949]{spi49}Spitzer L. and Schatzman E. 1949, ApJ 54, 195
 \bibitem[1951]{spi51}Spitzer L. and Tukey J. W. 1951, ApJ 114, 187S
\bibitem[1978]{spi78}Spitzer L. 1978, Physical processes in the interstellar medium, J. Wiley, N.Y
\bibitem[1932]{van}Van Vleck J. The theory of electric and magnetic susceptibilities 1932, Clarendon Press, Oxford
\bibitem[2012]{vos}Voshchinnikov  N. V. 2012, JQSRT 113, 2334
 \end{thebibliography}
\end{document}